\documentclass[aps,prb,twocolumn,superscriptaddress]{revtex4-2}

\usepackage{graphicx}

\usepackage{xcolor}
\usepackage{enumitem}
\usepackage[version=4]{mhchem}
\usepackage{gensymb}
\newcommand{\affekut}{\affiliation{%
Physikalisches Institut, Center for Quantum Science (CQ) and LISA$^+$,
Eberhard Karls Universit\"at T\"ubingen,
Auf der Morgenstelle 14,
72076 T\"ubingen, Germany}}

\newcommand{\affmpis}{\affiliation{%
Max Planck Institute for Solid State Research, 
Heisenbergstraße 1, 
70569 Stuttgart, Germany}}

\begin{document}

%\preprint{}

\title{Non-volatile multi-state electrothermal resistive switching in a strongly correlated insulator thin-film device}

%\thanks{}
\author{Farnaz Tahouni-Bonab}
\affekut
\author{Matthias Hepting}
\affmpis
\author{Theodor Luibrand}
\affekut
\author{Georg Cristiani}
\affmpis
\author{Christoph Schmid}
\affekut
\author{Gennady Logvenov}
\affmpis
\author{Bernhard Keimer}
\affmpis
\author{Reinhold Kleiner}
\affekut
\author{Dieter Koelle}
\affekut
\author{Stefan Guénon}
\affekut
\email{stefan.guenon@uni-tuebingen.de}
%

%\email{first.author@uni-tuebingen.de}
%\afflisa

\date{\today}

\begin{abstract}
Strongly correlated insulators, such as Mott or charge-transfer insulators, exhibit a strong temperature dependence in their resistivity.
Consequently, self-heating effects can lead to electrothermal instabilities in planar thin film devices of these materials.
When the electrical bias current exceeds a device-specific threshold, the device can switch from a high- to a low-resistance state through the formation of metallic filaments.
However, since the current and temperature redistribution effects that create these filaments are sustained by local Joule heating, a reduction of the bias current below a second (lower) threshold leads to the disappearance of filaments, and the device switches back into the high-resistance state.
Hence, electrothermal resistive switching is usually volatile.  
Here, on the contrary, we report on non-volatile resistive switching in a planar \ce{NdNiO3} thin-film device.
By combining electrical transport measurements with optical wide-field microscopy,
we provide evidence for a metallic filament that persists even after returning the bias current to zero.
We attribute this effect to the pronounced hysteresis between the cooling and heating branches in the resistance vs. temperature dependence of the device. 
At least one hundred intermediate resistance states can be prepared, which are persistent as long as the base temperature is kept constant.
Further, the switching process is non-destructive, and thermal cycling can reset the device to its pristine state. 
\end{abstract}

%\keywords{}

\maketitle

\section{Introduction}
The emerging field of neuromorphic computing involves the research and development of hardware and computing paradigms
that can effectively mimic the essentials of biological neural networks \cite{mead1989, schuller2015, christensen2022, ziegler2024}.
Resistive switching devices (RSDs) are particularly relevant because these simple two-terminal devices
can emulate two central neuromorphic features due to their memristive behavior \cite{ielmini2018,ielmini2018a, ziegler2018, ran2023, zhou2022}.
First, with volatile RSDs, it is possible to mimic the threshold behavior of neurons
\cite{pickett2012, ignatov2015, stoliar2017, yi2018, delvalle2020},
and second, with non-volatile RSDs, it is possible to emulate the information storage capability of synapses \cite{ziegler2015,hansen2017,burr2017, jeong2018,ielmini2018}.
RSDs change their intrinsic resistance from high to low if their electrical bias exceeds a device-specific threshold.
While volatile RSDs restore the high-resistance state at low electrical bias, the resistance change in non-volatile RSDs is persistent.
Various material systems showing different resistive switching effects are currently investigated \cite{valle2018, dittmann2021, hoffmann2022, goteti2021}.
The present work concerns strongly correlated insulator thin-film (SCITF) devices, which are gaining increasing attention in this research context. 
These materials, which show a pronounced increase in resistivity at low temperatures due to electron-electron interaction, include
Mott \cite{imada1998} and charge-transfer insulators \cite{catalano2018, bisogni2016, Johnston2014}.
SCTIF devices have a unique feature: the temperature dependence of the resistivity causes electrothermal instability,
which can trigger resistive switching due to the formation of a metallic filament
\cite{berglund1969, zimmers2013, kumar2013, guenon2013, manca2015, lange2021, rocco2022, luibrand2023}.
Because the electrothermal filament is sustained by local Joule heating, it should disappear at small bias currents.
Hence, electrothermal resistive switching in SCITF devices is expected to be volatile. 
Indeed, most studies on planar SCITF devices demonstrate volatile resistive switching \cite{valle2019, lange2021, 
adda2022, das2023, luibrand2023}.
Here, we report non-volatile resistive switching in a planar two-terminal device based on \ce{NdNiO3} (NNO), which is a charge transfer insulator that exhibits a concomitant metal-to-insulator, structural, and magnetic phase transition \cite{catalan2008, lu2016}.
In NNO thin films, the transition temperature is highly sensitive to the degree of epitaxial strain between film and substrate as well as the crystallographic orientation of the substrate surface \cite{liu2013a, catalano2015, suyolcu2021}. 
By combining electrical transport measurements with optical microscopy, we demonstrate that the non-volatile resistive switching is accompanied by the formation of a metallic filament that persists even at zero currents. 
We argue that the filament formation process is nevertheless electrothermal and that the filament persists due to the pronounced hysteresis in the resistance vs. temperature relation of the charge-transfer insulator device.  
%

%%%%%%%%%%%%%%%%%%%%%%%%%%%%%%%%%%%%%%%%%%%%%%%%%%%%%%%%%%%%%
% Figure 1 
% 
%%%%%%%%%%%%%%%%%%%%%%%%%%%%%%%%%%%%%%%%%%%%%%%%%%%%%%%%%%%%/i%
\begin{figure}
    \includegraphics[width=0.9\linewidth]{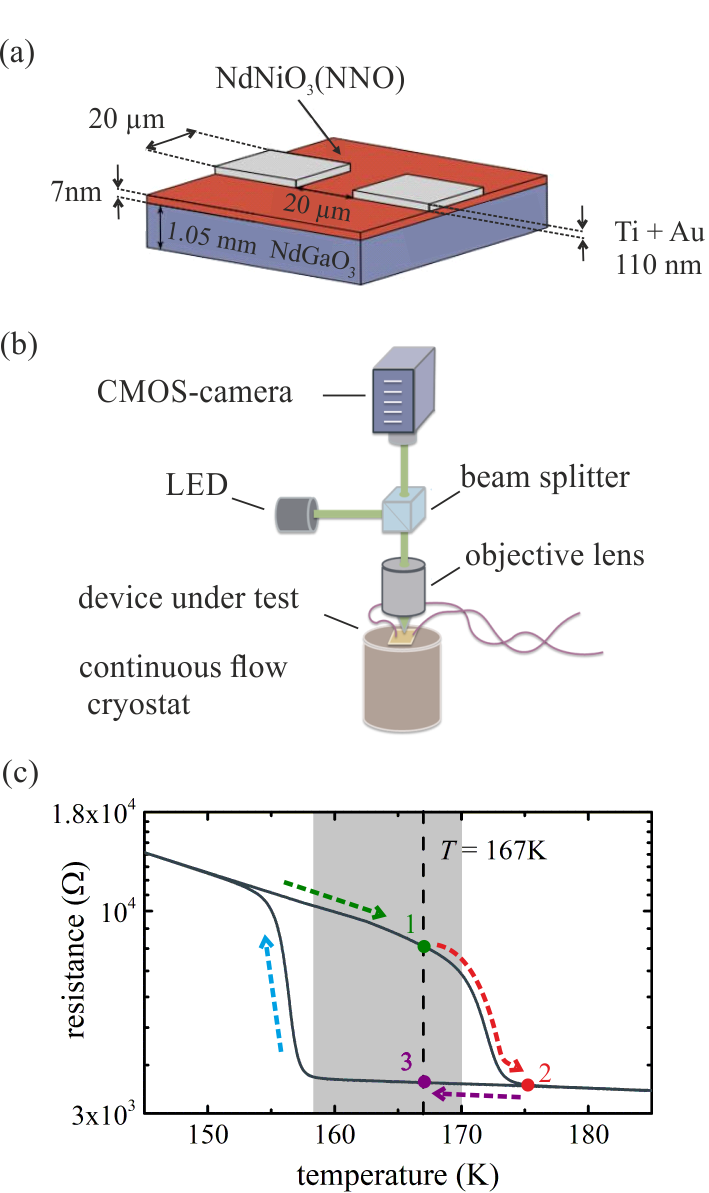}
    \caption{\label{fig:D}
        (a) Schematic of the two-terminal planar device with metal 
        electrodes on top of a continuous \ce{NdNiO3} (NNO) thin film. 
        (b) Schematic of the wide-field microscopy setup. 
        (c) Resistance vs. temperature relation of the DUT acquired with a current $I=100\,\mu$A during thermal cycling.
        The arrows indicate the sweep direction. 
        The cooling branch is shifted from the heating branch to lower temperatures by approximately 15\,K.
        For the electrical transport measurements, the device is put into a well defined state by interrupting 
        thermal cycling at the targeted base temperature of 167\,K indicated by a vertical dashed line. 
        This procedure leaves the device in an insulating state on the heating branch (point 1).
        Heating the device to a temperature above the insulator-to-metal transition (point 2) and cooling it to the original base temperature 
        puts it into a low-resistance, metallic state (point 3).
    }
\end{figure}
\section{Sample and Method}
%%%%%%%%%%%%%%%%%%%%%%%%%%%%%%%%%%%%%%%%%%%%%%%%%%%%%%%%%%%%%
% sample description 
% 
%%%%%%%%%%%%%%%%%%%%%%%%%%%%%%%%%%%%%%%%%%%%%%%%%%%%%%%%%%%%%
%
The sample under investigation consists of planar, two-electrode NNO thin-film devices.
Figure~\ref{fig:D}(a) shows a schematic of the device under test (DUT) with its dimensions indicated.
The 7\,nm thick NNO film was grown epitaxially via pulsed laser deposition from a stoichiometric NNO target on a (110)-oriented \ce{NdGaO3} single crystalline substrate.
The deposition temperature was 730\,\degree C, the oxygen partial pressure was 0.5\,mbar, and 
the \ce{KrF} excimer laser pulse rate was 2\,Hz with a 1.6\,J\,cm$^{-2}$ energy density.
The film was annealed in a 1\,bar oxygen atmosphere at 690\,\degree C for 30\,min.
As a result, the films show a sharp metal-to-insulator transition with hysteretic behavior between approximately 155 and 175\,K upon cooling and warming \cite{post2018}.
See the methods section and the supplementary information of \cite{post2018} for more details on the growth and physical properties of NNO thin films prepared by this growth procedure.
The metal electrodes on the continuous NNO film were structured via optical lithography and a lift-off process.
The \ce{Ti} buffer layer and the \ce{Au} layer were deposited by electron-beam evaporation with a total thickness of 110\,nm. 
%

%%%%%%%%%%%%%%%%%%%%%%%%%%%%%%%%%%%%%%%%%%%%%%%%%%%%%%%%%%%%%
% methods 
% 
%%%%%%%%%%%%%%%%%%%%%%%%%%%%%%%%%%%%%%%%%%%%%%%%%%%%%%%%%%%%%
%
This study used the wide-field mode of a high-resolution polarizing microscopy setup described in detail in \cite{lange2017, lange2018}.
A schematic of the setup is depicted in Fig.~\ref{fig:D}(b).
The sample is mounted in vacuum on the cold finger of liquid helium continuous flow cryostat with a temperature range of 4.2 to 300\,K.
The system facilitates simultaneous electrical transport measurements and imaging via wide-field microscopy.
We used a Keithley 2400 Sourcemeter for the electrical transport measurements as a current source.
The spatial resolution of the microscope is 0.48\,$\mu$m with a 532\,nm monochromatic LED illumination.
\section{Results}
%%%%%%%%%%%%%%%%%%%%%%%%%%%%%%%%%%%%%%%%%%%%%%%%%%%%%%%%%%%%%
% Figure 2 
% 
%%%%%%%%%%%%%%%%%%%%%%%%%%%%%%%%%%%%%%%%%%%%%%%%%%%%%%%%%%%%%
\begin{figure*}
    \includegraphics[width=0.98\linewidth]{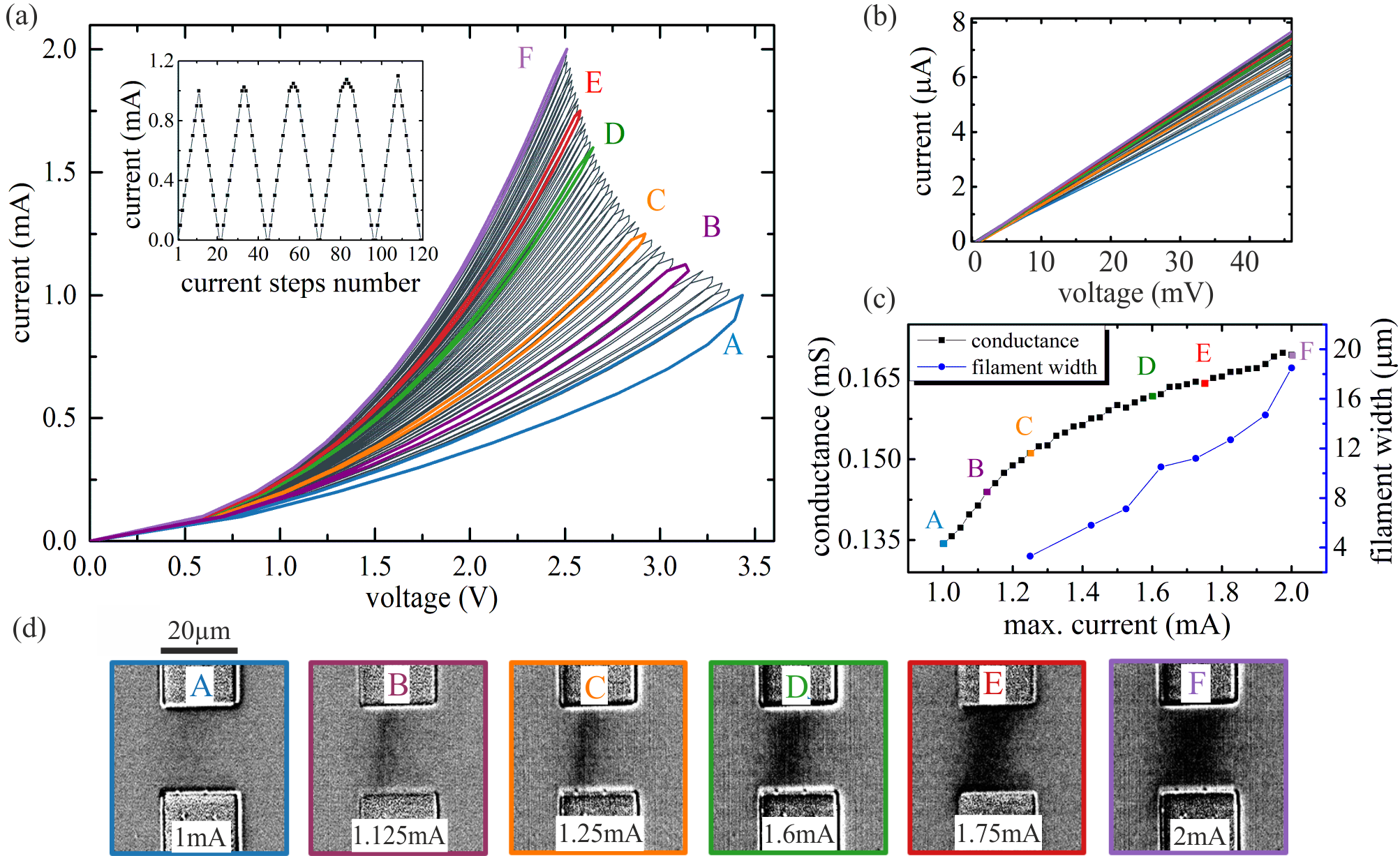}
    \caption{\label{fig:B}
        Non-volatile electrothermal resistive switching in a NNO two-terminal thin-film device initially prepared in the high-resistance 
        state at $T=167\,$K (point 1 in Fig.~\ref{fig:D}(c)). 
        (a) Current-voltage loops measured by applying triangular current sweeps. 
        The inset shows the first five sweeps, ramping $I$ from zero 
        to $I_{\rm max}$ and back in $100\,\mu$A steps with a 0.5\,s time-interval.
        The current interval was reduced to $25\,\mu$A for the highest current steps when 
        this was necessary to meet the targeted $I_{\rm max}$.
        $I_{\rm max}$ is consecutively increased from 1.0\,mA (loop A) in 
        $25\,\mu$A steps up to 2.0\,mA (loop F). 
        Note that the up-sweeps are almost congruent with the down-sweeps of the preceding loop, 
        except in the small section before reaching the maximum current.
        (b) Zoom in the origin of the graph shown in (a). 
        (c) Device conductance at a small bias current ($100\,\mu$A) vs. maximum current of 
        the preceding sweep. 
        The blue curve shows the corresponding filament width extracted from photomicrographs. 
        (d) Photomicrographs (differential images) of the DUT at zero current after the corresponding current-voltage loops have been acquired. 
        In B-F, the metallic filament is clearly visible. Bottom labels denote $I_\mathrm{max}$.   
}           
\end{figure*}

%%%%%%%%%%%%%%%%%%%%%%%%%%%%%%%%%%%%%%%%%%%%%%%%%%%%%%%%%%%%%
% describe the RT 
% 
%%%%%%%%%%%%%%%%%%%%%%%%%%%%%%%%%%%%%%%%%%%%%%%%%%%%%%%%%%%%%
%
Figure~\ref{fig:D}(c) shows the resistance $R$ vs. temperature  $T$ curve of the DUT acquired during thermal cycling.
This means that the sample was cooled from 200\,K to 80\,K and subsequently heated to 200\,K.
The strong resistance change between 158\,K and 154\,K in the cooling branch is caused by the metal-to-insulator transition (MIT) of the NNO thin film,
whereas the resistance change between 170\,K and 175\,K in the heating branch is due to the insulator-to-metal transition (IMT).
The hysteresis between the MIT and IMT indicates a first-order thermodynamic phase transition, 
and the steep $R(T)$ curves at the MIT and IMT suggest high film quality \cite{post2018}. 
Further, there is a broad temperature interval for which the device is entirely metallic during cooling and insulating during heating (indicated by the gray area in Fig.~\ref{fig:D}(c).
%
%Note that this feature cannot be found in the resistance vs. temperature curves in a study on volatile resistive switching in \ce{V2O3} \cite{lange2021}.
%
The hysteretic $R(T)$ behavior implies that thermal history can significantly impact the electrical transport properties. 
%

%%%%%%%%%%%%%%%%%%%%%%%%%%%%%%%%%%%%%%%%%%%%%%%%%%%%%%%%%%%%%
% describe non-volatile resistive switching in fB
% 
%%%%%%%%%%%%%%%%%%%%%%%%%%%%%%%%%%%%%%%%%%%%%%%%%%%%%%%%%%%%%
%
To demonstrate non-volatile, multi-state resistive switching, we put the DUT into a well-defined insulating state in the heating branch by interrupting the thermal cycle at 167\,K (see Fig.~\ref{fig:D}(c)).
Then, we applied a sequence of triangular current-sweeps, ramping the current $I$ from zero to a maximum current $I_{\rm max}$ and back to zero (see inset of Fig.~\ref{fig:B}(a)). 
Within this sequence $I_{\rm max}$ was first set to 1\,mA and subsequently increased in $25\,\mu$A steps up to 2\,mA. 
Figure~\ref{fig:B}(a) shows the current $I$ vs. voltage $V$ curves recorded throughout the entire sweep sequence.
The zoom of the $I(V)$ curves close to the origin in Fig.~\ref{fig:B}(b) demonstrates that even at currents very close to zero, the resistance after each cycle does not return to its initial state before the cycle started, i.e., the resistive switching is non-volatile. 
Moreover, with increasing number of sweeps (increasing $I_{\rm max}$), the conductance (slope of the $I(V)$ curves) monotonically increases.    
To illustrate this more quantitatively, Fig.~\ref{fig:B}(c) shows the dependence of the conductance $I/V$ at $I=100\,\mu$A as a function of $I_{\rm max}$ recorded on the return branch of each current sweep. 
This observation clearly shows that we can create multiple states by choosing different values for $I_{\rm max}$. 
To investigate the observed non-volatile, multi-state resistive switching further, we recorded photomicrographs of the DUT with the wide-field microscope at the end of every single current sweep, i.e., after returning to zero bias current.
Figure \ref{fig:B}(d) shows a selection of images, recorded for the $I(V)$ loops that are labeled from A to F in Fig.~\ref{fig:B}(a). 
Note that these are differential images with the image of the pristine device (after preparing the DUT in point 1 in Fig.~\ref{fig:D}(c) and before starting the sequence of current sweeps) subtracted to enhance contrast.
The rectangles at the top and bottom are the metal electrodes of the two-terminal device.
In the photomicrographs, metallic and insulating regions of the NNO thin film are clearly distinguishable due to their different reflectivity, with the metallic areas appearing darker than the insulating ones \cite{luibrand2023}.
Clearly, Fig.~\ref{fig:B}(d) illustrates that a metallic (dark) filament connects the two electrodes and we find that even after returning the current to zero, this filament persists. 
Moreover, with increasing $I_{\rm max}$ the width of the filament increases. 
To quantify this, we evaluated the filament width from the optical images, and added the filament width vs. $I_{\rm max}$ to Fig.~\ref{fig:B}(c).
Notably, the monotonic increase in conductance correlates with a monotonic increase in filament width. 
The series of images shown in Fig.~\ref{fig:B}(d) can explain the observed non-volatile, multi-state resistive switching:
We find that even after returning to $I=0$ a metallic filament persists, which causes a slight increase in the overall conductance of the DUT, and the observed increase in filament width with increasing $I_{\rm max}$ accounts for the increase in conductance with increasing $I_{\rm max}$, generating multiple resistive states. 
%

\begin{figure}
    \includegraphics[width=0.8\linewidth]{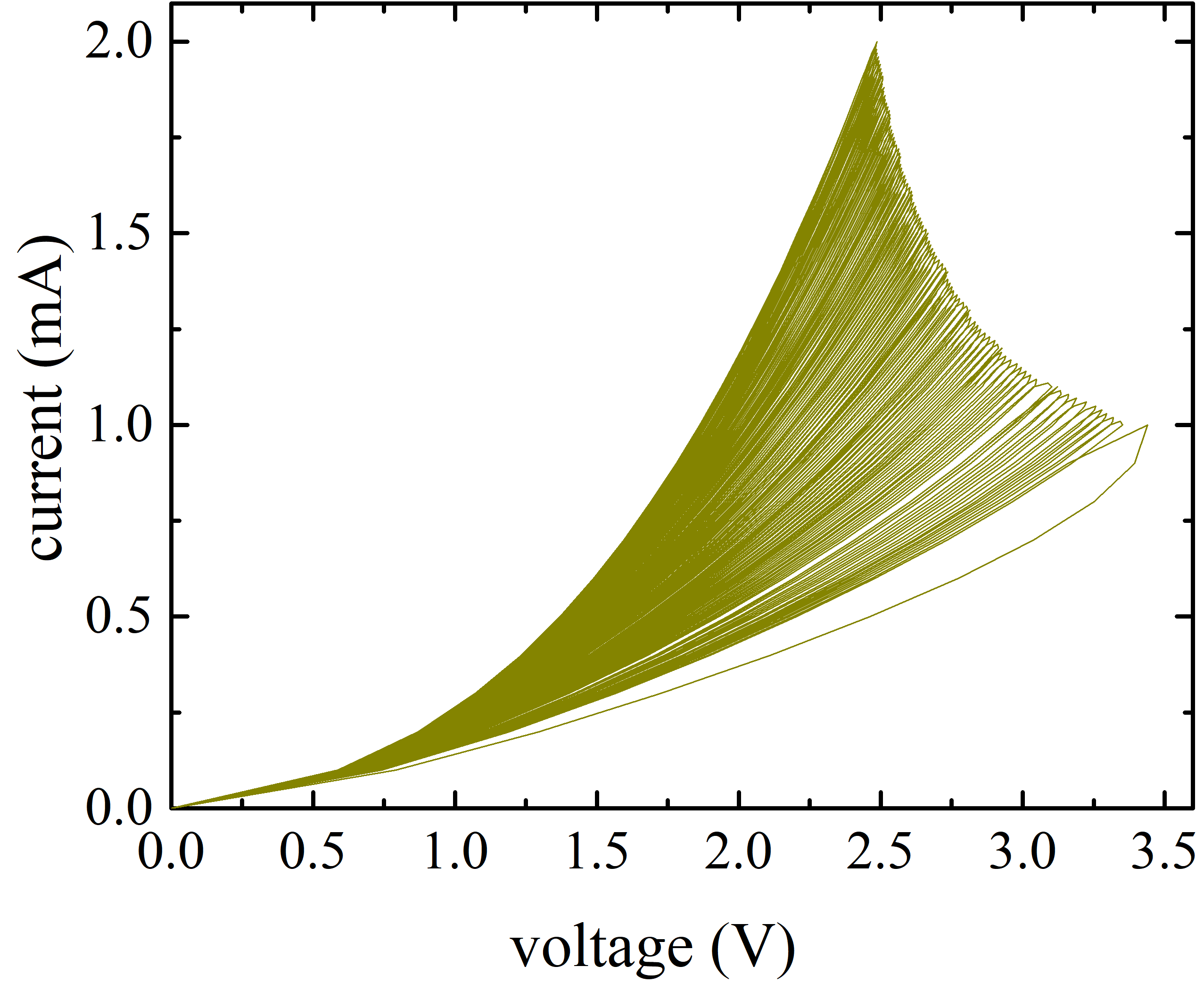}
    \caption{\label{fig:E}
        Series of current-voltage loops, analog to Fig.~\ref{fig:B}(a), demonstrating the information storage capability of the device. 
        $I_{\rm max}$ was increased in $10\,\mu$A steps. 
    }
\end{figure}

%%%%%%%%%%%%%%%%%%%%%%%%%%%%%%%%%%%%%%%%%%%%%%%%%%%%%%%%%%%%%
% describe non-volatile resistive switching in fB
% 
%%%%%%%%%%%%%%%%%%%%%%%%%%%%%%%%%%%%%%%%%%%%%%%%%%%%%%%%%%%%%
%
In addition, we repeated the sweep sequence with a smaller step size of $10\,\mu$A for $I_{\rm max}$ (Fig.~\ref{fig:E}).
As shown, 100 intermediate states can be prepared.
This suggests that storing information in an analog state with high resolution in the device might be possible.
Further, the intermediate states are stable if the base temperature remains unchanged whereas these non-volatile states can be erased by thermal cycling allowing to restore the initial state of the device.
%

%%%%%%%%%%%%%%%%%%%%%%%%%%%%%%%%%%%%%%%%%%%%%%%%%%%%%%%%%%%%%%
% Figure 3
% 
%%%%%%%%%%%%%%%%%%%%%%%%%%%%%%%%%%%%%%%%%%%%%%%%%%%%%%%%%%%%%

\section{Discussion}
The observed non-volatile resistive switching is primarily due to a persistent metallic filament caused by the hysteretic $R(T)$, in which the IMT- and MIT-temperature intervals are separated due to the steep (almost vertical) transitions between the high-resistance insulating and the low-resistance conducting states, as discussed below.

Let us first discuss, the concept of electrothermal resistive switching in a planar SCITF device with a non-hysteretic $R(T)$.
As explained in Appendix \ref{sec:app_rs}, electrothermal resistive switching is driven by Joule heating.
The steep decline of the $R(T)$ due to the IMT acts as a feedback mechanism, which can lead to electrothermal instability.
At currents above a certain threshold,  a homogeneous current distribution becomes unstable because small perturbations in the current density are amplified by the local increase of Joule heating that reduces the electrical resistance of the SCITF.
This feedback mechanism funnels more and more current into a small section, causing resistive switching.
At the end of the resistive switching event, a metallic filament is formed between the two electrodes of the device, reducing its resistance.
The electrical current flows almost entirely through the metallic filament, which creates a considerable amount of Joule heating confined in a small area.
The local excess temperature caused by the Joule heating is high enough to keep the SCITF above the IMT temperature in the metallic state.
In this manner, Joule heating sustains and stabilizes the metallic filament.
If the current is increased further after the device has switched to the low-resistance state, the Joule heating in the filament is increased, and as a result, the filament becomes wider.
Equivalently,  the filament width scales down when the current is reduced.
If the electrical current is reduced below a second threshold value (smaller than the threshold value for electrothermal resistive switching from the high-resistance to the low-resistance state), Joule heating becomes too weak to support the metallic filament, and consequently, the device switches back to the high-resistance state with a homogeneous current distribution. 
Thus, electrothermal resistive switching is generally volatile.
For an illustration of volatile resistive switching in a \ce{V2O3} device, see e.g. Ref.~\cite{lange2021} with VIDEO 1, therein. 

But why do we observe non-volatile resistive switching in an NNO device?
The high crystalline quality and phase purity of the NNO film lead to very steep changes of the resistance across the IMT and MIT, while the hysteretic temperature regime is broad, with an interval of at least 10\,K, in which the thin film is in a high-resistivity insulating state during heating and a low-resistivity metallic state during cooling. 

Note that the base temperature for the electrical transport measurements (Fig.~\ref{fig:D}(c))
lies precisely in this interval. 
If the electrical current is increased after the device was brought into a pristine high-resistance state by thermal cycling (indicated by point 1 in Fig.~\ref{fig:D}(c)), the situation is the same as described two paragraphs above, and the homogeneous current distribution becomes unstable, which results in the Joule-heating-driven formation of a metallic filament.
However, the situation is distinctively different if the current is reduced after a metallic filament was created.
Because the NNO thin-film areas, which are driven into the metallic state via Joule heating, cannot return to the insulating state (consider the thermal sequence 1 to 3 in Fig.~\ref{fig:D}(c)), the filament cannot shrink and persists even at zero bias current.
Hence, the NNO device stays in the low-resistance state, and the resistive switching is non-volatile. 

Next, we address the case when the electrical current is increased from zero bias after a metallic filament already exists.
Most of the current is confined to the filament.
For small currents, the Joule heating is insufficient to raise the local temperature above the threshold of the IMT, and the filament does not change.
Consequently, the up-sweep IV curves in Fig.~\ref{fig:B}(a) are almost congruent with the down-sweep curves of the preceding loops. 
However, suppose the current is increased above the maximum value that was applied to form the filament before.
In that case, the excess temperature created by Joule heating in its vicinity is high enough to metallize additional NNO film areas, and the filament width increases.
With this mechanism, intermediate resistance states can be prepared (see Figs.~\ref{fig:B} and \ref{fig:E}). 
Compared to the idealized case of volatile electrothermal resistive switching discussed in Appendix \ref{sec:app_rs}, with a single abrupt resistive switching event followed by a vertical progression of the IV curve (Fig.~\ref{fig:C}(e)), for the non-volatile switching case describe here, the switching seems gradual, and considering the envelope connecting $I_{\rm max}$ of all the loops, the progression is diagonal (see Fig.~\ref{fig:B}(a)).
As shown in Appendix \ref{sec:rs_diff_T}, it turns out that abrupt resistive switching can also occur at devices discussed here, however only at slightly lower temperatures, closer to the MIT. Moreover, the gradual switching observed at higher temperatures is also discussed in Appendix \ref{sec:rs_diff_T} in more detail and might be attributed to boundary effects or thermal characteristics of the NNO thin film. 
Because the electrode width limits the maximum filament width, increasing the dynamic range of the non-volatile resistive switching should be possible by using a device layout with a larger aspect ratio between the electrode width and the gap size.
\section{Conclusion}
In conclusion, we have demonstrated non-volatile electrothermal resistive switching due to the formation of a persistent metallic filament.
Unlike an ion-migration forming process in electronic oxides resistive switching devices, the filament is created due to electrothermal instability.
The filament configuration is spatially fixed due to the distinct separation of the heating and cooling branches in the
resistance vs. temperature relation.
We have demonstrated that the device can be prepared in at least one hundred states with different resistances, 
which are stable as long as the temperature is kept constant. 
The original pristine state can be restored by thermal cycling.
Although the non-volatile resistive switching is, in this sense, not permanent, electrothermal resistive switching devices offer considerably higher endurance than devices based on ion displacement, which makes them attractive candidates for applications where information needs to be stored and erased at high frequencies.
The results can be applied to devices with similar resistance vs. temperature relations, like \ce{VO2} devices, because the non-volatile behavior is caused by $R(T)$-curve characteristics.
Our report on a non-volatile resistive switching device, which is based on electrothermal instability,
not only underscores the potential of SCITF devices but also opens up exciting possibilities
for their application in neuromorphic systems, e.g., using their information storage capacity for recording synaptic weights. 
Several preceding studies on SCITF-based resistive switching devices reported non-volatile effects
\cite{rozenberg2004, driscoll2009, coy2010, ha2011, hu2011, pellegrino2012, cai2013, shi2013, seo2014, patel2020, shabalin2020, ma2021, ma2023a}, but the critical connection to electrothermal filament formation has remained elusive. Therefore, this work is a significant step forward in understanding these devices and their potential impact on the emerging research field of neuromorphic computing.
\section{Acknowledgments}
We thank the instrument scientists Ronny L\"offler and Markus Turad of the research and service facility LISA$^+$ for their support in patterning the sample. 
F.T. acknowledges support from the Landesgraduiertenförderung Baden-Württemberg. 
T.L acknowledges support from the Cusanuswerk, Bischöfliche Studienförderung.

\appendix
\section{Volatile Electrothermal Resistive Switching}
\label{sec:app_rs}
%%%%%%%%%%%%%%%%%%%%%%%%%%%%%%%%%%%%%%%%%%%%%%%%%%%%%%%%%%%%%%
% Figure 4
% 
%%%%%%%%%%%%%%%%%%%%%%%%%%%%%%%%%%%%%%%%%%%%%%%%%%%%%%%%%%%%%
\begin{figure*}
    \includegraphics[width=0.98\linewidth]{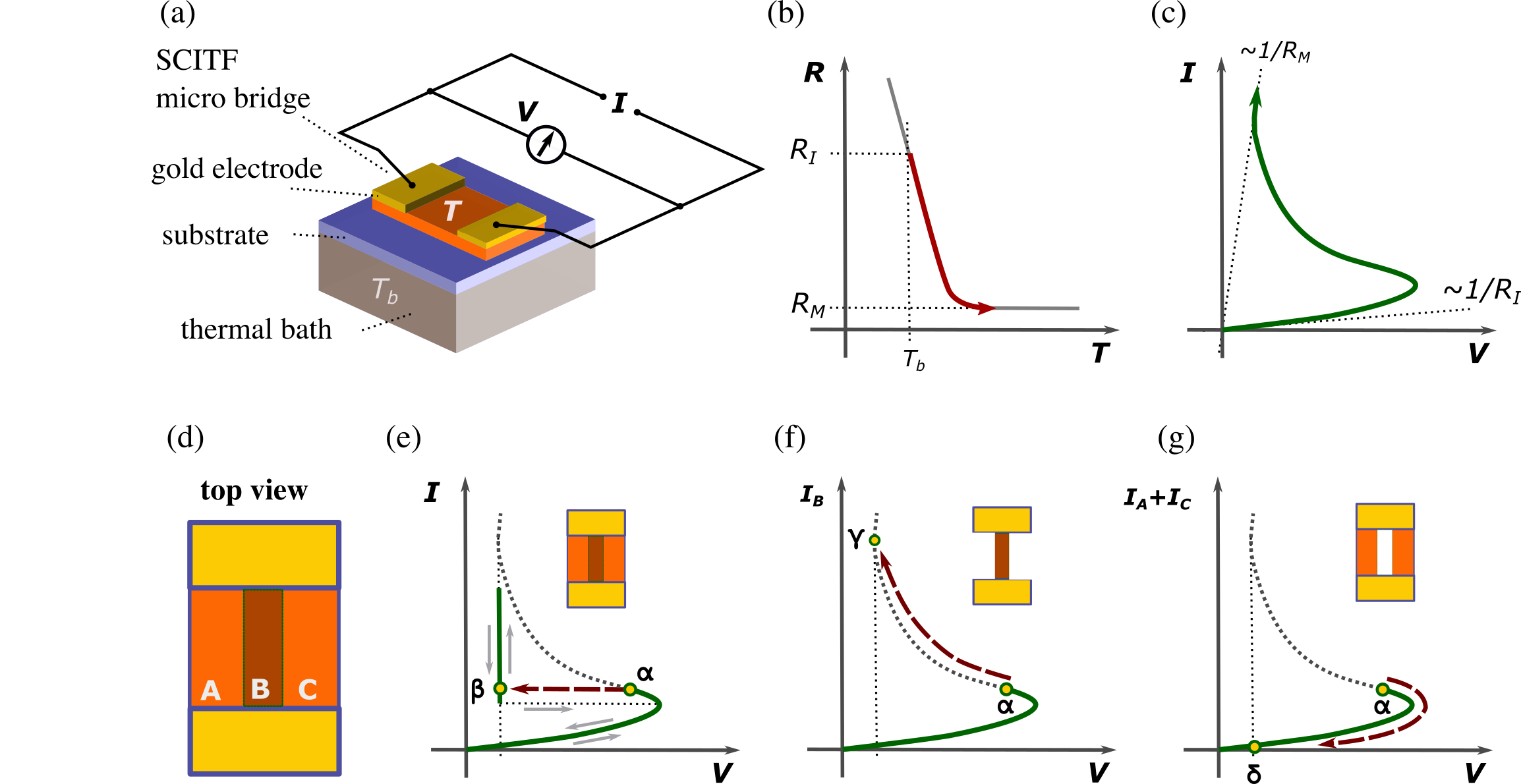}
    \caption{\label{fig:C}
    Electrothermal instability causing volatile resistive switching of a strongly correlated insulator thin-film (SCITF) microbridge.  (a) Schematic of the device. $T$ device temperature and $T_b$ base temperature. (b) Resistance vs. temperature dependence. $R_I$ and $R_M$ are the device resistance in the insulating and metallic state. The red curve indicates the evolution of the device resistance due to Joule heating under the assumption of a homogenous temperature distribution (no edge cooling effects) (c) The current-voltage relation is S-shaped due to Joule heating. (d) The microbridge is subdivided into three sections (A, B, and C) for a thought experiment. (e) Current-voltage relation of the whole device. Light gray arrows indicate sweep directions. At bias point $\alpha$, in the negative differential resistance regime, the device becomes electrothermally unstable and switches to bias point $\beta$ (red-dashed line). (f) Current-voltage relation of section B. Section B switches to bias point $\gamma$ with a positive differential resistance and a high current. (g) Current-voltage relation of the sum of the currents in sections A and C. Sections A and C switch to bias point $\delta$ with a positive differential resistance and a low current. Note that the scales of the voltage axes in (e-g) are the same, while the scales of the current axes vary due to different form factors.}
\end{figure*}
%%%%%%%
The phenomenon of electrothermal instability is created by a drastic change in the resistance vs. temperature dependence of an electronic material.
It is widespread and observed in many electrical devices like superconducting wires \cite{gurevich1987}, hot electron bolometers \cite{doenitz2007} or superconducting THz-radiation emitters \cite{gross2012}.

According to M. Büttiker and R. Landauer \cite{buttiker1982} one of the first researchers who captured the concept 
of electrothermal instability was E. Spenke \cite{spenke1936,spenke1936a}. B.K. 
Ridley published a more recent study on that topic \cite{ridley1963}.
In the following E. Spenke's main ideas are applied to electrothermal resistive switching in 
SCITF-devices.
Figure~\ref{fig:C}(a) shows a planar thin-film device schematic. A microbridge design is chosen 
for the sake of simplicity. 
Figure~\ref{fig:C}(b) depicts the corresponding resistance vs. 
temperature relation. Due to the IMT, there is a steep decline in the heating branch. 
How does current-induced self-heating affect the current-voltage relation of the device? We 
assume a bath temperature $T_b$ for which the device is insulating (Fig.~\ref{fig:C}(b)). 
If we apply a current using a constant current source, the device temperature will increase due to Joule heating. 
We neglect boundary effects and assume that the Joule heating induces a homogeneous rise in device temperature $T$. 
In this case, the Joule-heating-induced resistance change will follow the resistance vs.
temperature curve in Fig.~\ref{fig:C}(b).
Consider Fig.~\ref{fig:C}(c) for a qualitative discussion of the current-voltage relation. 
The Joule heating effect is negligible at small currents, and the device resistance is $R_I$. 
This results in a linear current-voltage-curve progression with a slope proportional to $1/R_I$. 
If the current is increased further, Joule heating affects the device temperature, 
which reduces the device resistance, which causes an up-bending of the $I(V)$ curve.
With further increasing current, Joule heating and the resulting bending of the $I(V)$ curve is so intense that it 
enters a regime with negative differential resistance.
Suppose the Joule heating is so intense that the device is heated into the metallic state. 
In that case, the self-heating of the device no longer affects its resistance because the resistance-temperature 
curve is almost horizontal ($R(T)=R_M$ constant) in the metallic regime, and the current-voltage-curve progresses with a slope proportional to $1/R_M$.
Hence, assuming a homogeneous current and temperature distribution, 
self-heating creates an S-shaped current-voltage relation.
However, in the current bias regime for which the current-voltage relation has a negative 
differential resistance, the device is unstable, and an electrothermal resistive switching can happen due to current and temperature redistribution, as will be explained in the following thought experiment.
Let us divide the device into three sections (see Fig.~\ref{fig:C}(d)). 
All the sections connect the electrodes with two sides parallel to the current direction. 
Hence, they can also be considered as microbridges. 
Let us assume the device is at the bias point $\alpha$ in the negative differential 
resistance regime. 
Due to the electrothermal instability, the following can happen: 
section B jumps to bias point $\gamma$ (Fig.~\ref{fig:C}(f)) with a high current and positive differential 
resistance and to compensate section A and C jump to bias point $\delta$ (Fig.~\ref{fig:C}(g)) 
with a low current and positive differential resistance as well.
Because the sample is biased with a current source, the $I(V)$ curve has to follow a horizontal load line, 
and the width of section B has to be such that the whole device jumps to bias point $\beta$ (Fig.~\ref{fig:C}(e)).
Note, that due to the one-to-one relationships between temperature, resistance and current 
discussed above (Fig.~\ref{fig:C}(b and d)) section B at bias point $\gamma$ has to be at an 
elevated temperature in the metallic state. 
Hence, the electrothermal resistive switching is accompanied by the formation of a metallic filament
and is the result of a temperature and current redistribution effect. 
If the current increases after the switching event, the voltage over the device 
stays constant, and the width of section B (metallic filament) increases. 
Likewise, the width of the filament reduces if the current is reduced, and for small currents, 
the filament will disappear because the Joule heating is too small to support it. 
Hence, the electrothermal resistive switching is volatile because the device 
returns to its high-resistance state for small currents.
An excellent example of volatile resistive switching can be found in a study on \ce{V2O3}-devices 
\cite{lange2021}.
In this publication, an animation presenting the filament formation, the vertical progression of the current-voltage relation with the filament growth and shrinkage, and the disappearance of the filament can be found in VIDEO 1.
\section{Resistive Switching at Different Base Temperatures}
\label{sec:rs_diff_T}
\begin{figure}
    \includegraphics[width=0.9\linewidth]{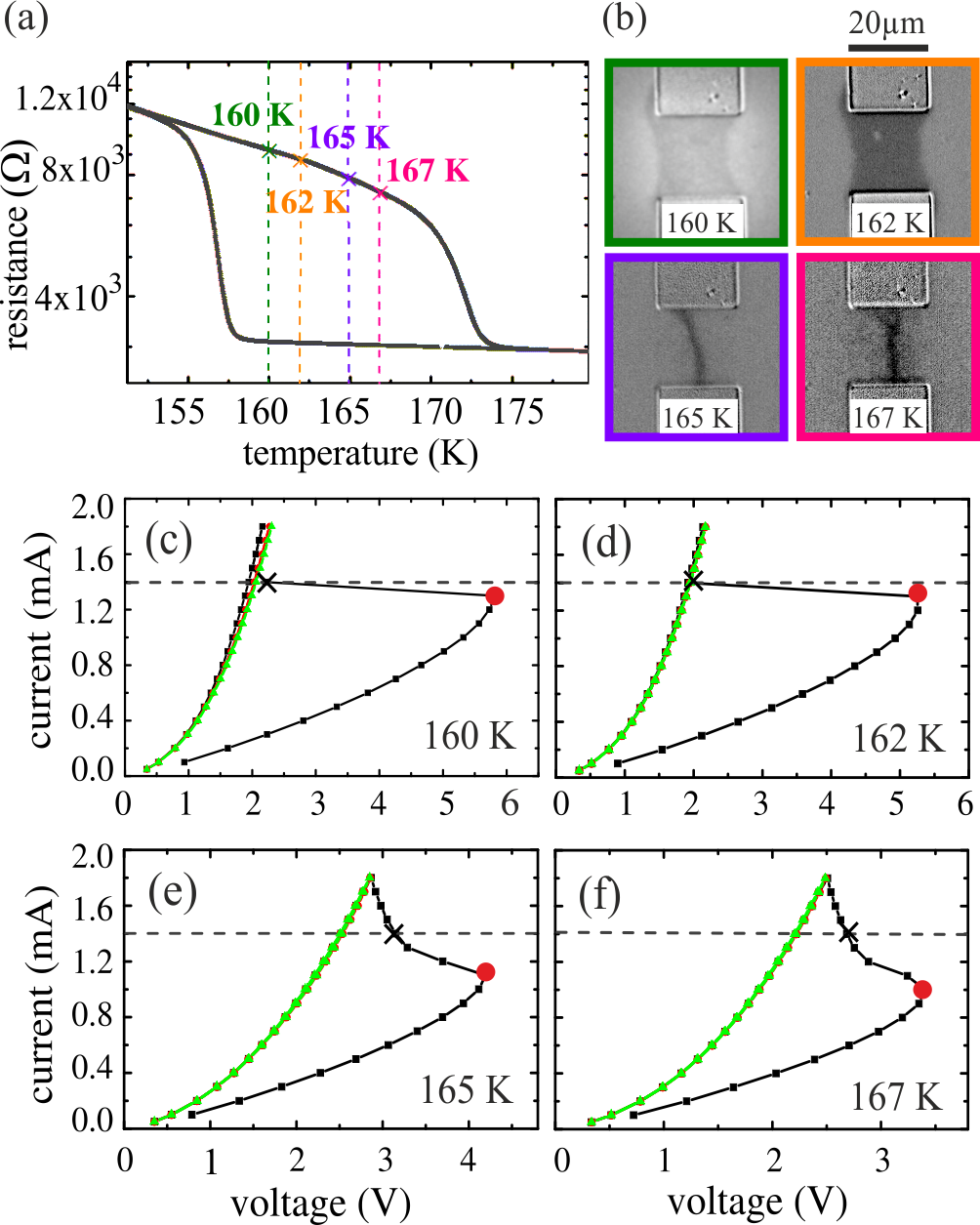}
    \caption{\label{fig:A}
        Non-volatile resistive switching at different base temperatures in an interval above the MIT and below the IMT. (a) $R(T)$ curves (cooling and heating branches) with $I=100\,\mu$A. The four base temperatures are indicated by vertical dashed lines. (b) Photomicrographs acquired at 1.4\,mA bias current (the bias points during imaging are indicated by black crosses in subfigures (c-f)) after the device has switched into a low-resistance state. (c-f) Current vs. voltage curves acquired during three consecutive current sweeps from zero to 1.8\,mA bias current and back to zero after the device was brought into a pristine state via thermal cycling. Black curve: first current sweep. Red curve: second current sweep. Green curve: third current sweep. Red dots indicate the bias points at the breakdown voltage.}
\end{figure}
In addition to studying resistive switching at the base temperature of 167\,K (as discussed in the main text),  we investigated resistive switching at lower temperatures by applying three consecutive current sweeps (at each base temperature) after the sample was brought into a pristine state via thermal cycling.
All base temperatures fall in the bistable interval between the MIT and IMT (Fig.~\ref{fig:C}(a)).
Figure \ref{fig:C}(b) depicts photomicrographs of the device at a bias current of 1.4\,mA during the increasing current sweep of the first loop.
All images are differential images, i.e., the image of the pristine device was subtracted to enhance contrast. The current vs. voltage curves are plotted in Fig.~\ref{fig:C}(c-f). 
The IV curves acquired at the two lower temperatures (160 and 166\,K) both show abrupt resistive switching in the first loop, indicated by an almost horizontal line.
The up- and down-sweep curves of the second and third loop lie precisely on top of the down-sweep curve of the first loop.
Hence, the resistive switching is non-volatile.
The corresponding photomicrographs in Fig.~\ref{fig:C}(b) that were acquired immediately after the resistive switching event show a broad metallic filament extending almost over the whole area between the electrodes.  
On the contrary, the current-voltage curves acquired at the two higher temperatures (165 and 167\,K) do not show abrupt but gradual switching in the up-sweep of the first loop.
Again, the down-sweep of the first loop and the up- and down-sweeps of the subsequent loops are congruent, indicating non-volatile resistive switching.
The corresponding photomicrographs in Fig.~\ref{fig:C}(b) show that the filaments are considerably narrower than the electrodes. 
To explain the differences between resistive switching at different base temperatures, consider the bias point in the up-sweep of the first loop at the breakdown voltage (red dot in the IV curve), i.e., right before the first resistive switching event.
First, note that the breakdown voltage becomes larger with lower base temperatures.
Second, the current required to trigger resistive switching is noticeably increased in case of the two lower base temperatures.
Combining these two observations, we can conclude that decreasing temperature increases the electric power required to cause resistive switching.
Considering the discussion of electrothermal resistive switching in Appendix \ref{sec:app_rs}, this makes sense.
Since, at low temperatures, the excess temperature to reach the metallic state is large, more Joule heating is needed to bring the device into the electrical bias regime where the system becomes electrothermally unstable.
Because a rather high electric current is needed to cause resistive switching at low temperatures, the filament width immediately after the switching event is quite large, so it can support this high current. 
Regarding the gradual resistive switching observed at the two higher base temperatures ($T=165\,$K and $T=167\,$K), such an IV shape in an electrothermal unstable device is not new and was, for instance, observed during electrothermal switching (hot spot formation) in BSCCO-THz emitters \cite{gross2012}, where it can be attributed to the fact that the three-dimensional character and the boundaries of the system affect temperature redistribution.
Similar effects could be at play in the case of the NNO device at base temperatures near the IMT.
For instance, the metallic electrodes of the two-terminal device provide thermal cross-coupling, affecting the temperature distribution.
Another explanation for the gradual resistive switching might be that the current and temperature redistribution mechanism responsible for filament formation is impeded by the creation of isolated metallic domains (smaller than the spatial resolution of the microscope) when the device is electrically biased for the first time after thermal cycling.
This creates a network of percolation paths between the electrodes, which prevents the electrical current from being funneled exclusively in a small section of the device.
Therefore, the resistive switching is gradual during the first current up-sweep. This might also explain why the filament cannot be observed in the photomicrograph acquired after loop A (Fig.~\ref{fig:B}(d)).
\bibliography{FTb-ms1-bibliography}

\begin{thebibliography}{65}%
\makeatletter
\providecommand \@ifxundefined [1]{%
 \@ifx{#1\undefined}
}%
\providecommand \@ifnum [1]{%
 \ifnum #1\expandafter \@firstoftwo
 \else \expandafter \@secondoftwo
 \fi
}%
\providecommand \@ifx [1]{%
 \ifx #1\expandafter \@firstoftwo
 \else \expandafter \@secondoftwo
 \fi
}%
\providecommand \natexlab [1]{#1}%
\providecommand \enquote  [1]{``#1''}%
\providecommand \bibnamefont  [1]{#1}%
\providecommand \bibfnamefont [1]{#1}%
\providecommand \citenamefont [1]{#1}%
\providecommand \href@noop [0]{\@secondoftwo}%
\providecommand \href [0]{\begingroup \@sanitize@url \@href}%
\providecommand \@href[1]{\@@startlink{#1}\@@href}%
\providecommand \@@href[1]{\endgroup#1\@@endlink}%
\providecommand \@sanitize@url [0]{\catcode `\\12\catcode `\$12\catcode
  `\&12\catcode `\#12\catcode `\^12\catcode `\_12\catcode `\%12\relax}%
\providecommand \@@startlink[1]{}%
\providecommand \@@endlink[0]{}%
\providecommand \url  [0]{\begingroup\@sanitize@url \@url }%
\providecommand \@url [1]{\endgroup\@href {#1}{\urlprefix }}%
\providecommand \urlprefix  [0]{URL }%
\providecommand \Eprint [0]{\href }%
\providecommand \doibase [0]{https://doi.org/}%
\providecommand \selectlanguage [0]{\@gobble}%
\providecommand \bibinfo  [0]{\@secondoftwo}%
\providecommand \bibfield  [0]{\@secondoftwo}%
\providecommand \translation [1]{[#1]}%
\providecommand \BibitemOpen [0]{}%
\providecommand \bibitemStop [0]{}%
\providecommand \bibitemNoStop [0]{.\EOS\space}%
\providecommand \EOS [0]{\spacefactor3000\relax}%
\providecommand \BibitemShut  [1]{\csname bibitem#1\endcsname}%
\let\auto@bib@innerbib\@empty
%</preamble>
\bibitem [{\citenamefont {Mead}(1989)}]{mead1989}%
  \BibitemOpen
  \bibfield  {author} {\bibinfo {author} {\bibfnamefont {C.}~\bibnamefont
  {Mead}},\ }\href@noop {} {\emph {\bibinfo {title} {Analog {{VLSI}} and Neural
  Systems}}}\ (\bibinfo  {publisher} {Addison-Wesley Longman Publishing Co.,
  Inc.},\ \bibinfo {address} {USA},\ \bibinfo {year} {1989})\BibitemShut
  {NoStop}%
\bibitem [{\citenamefont {Schuller}\ \emph {et~al.}(2015)\citenamefont
  {Schuller}, \citenamefont {Stevens}, \citenamefont {Pino},\ and\
  \citenamefont {Pechan}}]{schuller2015}%
  \BibitemOpen
  \bibfield  {author} {\bibinfo {author} {\bibfnamefont {I.~K.}\ \bibnamefont
  {Schuller}}, \bibinfo {author} {\bibfnamefont {R.}~\bibnamefont {Stevens}},
  \bibinfo {author} {\bibfnamefont {R.}~\bibnamefont {Pino}},\ and\ \bibinfo
  {author} {\bibfnamefont {M.}~\bibnamefont {Pechan}},\ }\href@noop {} {\emph
  {\bibinfo {title} {Neuromorphic {{Computing}} -- {{From Materials Research}}
  to {{Systems Architecture Roundtable}}}}},\ \bibinfo {type} {Tech. Rep.}\
  (\bibinfo {year} {2015})\BibitemShut {NoStop}%
\bibitem [{\citenamefont {Christensen}\ \emph {et~al.}(2022)\citenamefont
  {Christensen}, \citenamefont {Dittmann}, \citenamefont {{Linares-Barranco}},
  \citenamefont {Sebastian}, \citenamefont {Gallo}, \citenamefont {Redaelli},
  \citenamefont {Slesazeck}, \citenamefont {Mikolajick}, \citenamefont {Spiga},
  \citenamefont {Menzel}, \citenamefont {Valov}, \citenamefont {Milano},
  \citenamefont {Ricciardi}, \citenamefont {Liang}, \citenamefont {Miao},
  \citenamefont {Lanza}, \citenamefont {Quill}, \citenamefont {Keene},
  \citenamefont {Salleo}, \citenamefont {Grollier}, \citenamefont
  {Markovi{\'c}}, \citenamefont {Mizrahi}, \citenamefont {Yao}, \citenamefont
  {Yang}, \citenamefont {Indiveri}, \citenamefont {Strachan}, \citenamefont
  {Datta}, \citenamefont {Vianello}, \citenamefont {Valentian}, \citenamefont
  {Feldmann}, \citenamefont {Li}, \citenamefont {Pernice}, \citenamefont
  {Bhaskaran}, \citenamefont {Furber}, \citenamefont {Neftci}, \citenamefont
  {Scherr}, \citenamefont {Maass}, \citenamefont {Ramaswamy}, \citenamefont
  {Tapson}, \citenamefont {Panda}, \citenamefont {Kim}, \citenamefont {Tanaka},
  \citenamefont {Thorpe}, \citenamefont {Bartolozzi}, \citenamefont {Cleland},
  \citenamefont {Posch}, \citenamefont {Liu}, \citenamefont {Panuccio},
  \citenamefont {Mahmud}, \citenamefont {Mazumder}, \citenamefont {Hosseini},
  \citenamefont {Mohsenin}, \citenamefont {Donati}, \citenamefont {Tolu},
  \citenamefont {Galeazzi}, \citenamefont {Christensen}, \citenamefont {Holm},
  \citenamefont {Ielmini},\ and\ \citenamefont {Pryds}}]{christensen2022}%
  \BibitemOpen
  \bibfield  {author} {\bibinfo {author} {\bibfnamefont {D.~V.}\ \bibnamefont
  {Christensen}}, \bibinfo {author} {\bibfnamefont {R.}~\bibnamefont
  {Dittmann}}, \bibinfo {author} {\bibfnamefont {B.}~\bibnamefont
  {{Linares-Barranco}}}, \bibinfo {author} {\bibfnamefont {A.}~\bibnamefont
  {Sebastian}}, \bibinfo {author} {\bibfnamefont {M.~L.}\ \bibnamefont
  {Gallo}}, \bibinfo {author} {\bibfnamefont {A.}~\bibnamefont {Redaelli}},
  \bibinfo {author} {\bibfnamefont {S.}~\bibnamefont {Slesazeck}}, \bibinfo
  {author} {\bibfnamefont {T.}~\bibnamefont {Mikolajick}}, \bibinfo {author}
  {\bibfnamefont {S.}~\bibnamefont {Spiga}}, \bibinfo {author} {\bibfnamefont
  {S.}~\bibnamefont {Menzel}}, \bibinfo {author} {\bibfnamefont
  {I.}~\bibnamefont {Valov}}, \bibinfo {author} {\bibfnamefont
  {G.}~\bibnamefont {Milano}}, \bibinfo {author} {\bibfnamefont
  {C.}~\bibnamefont {Ricciardi}}, \bibinfo {author} {\bibfnamefont {S.-J.}\
  \bibnamefont {Liang}}, \bibinfo {author} {\bibfnamefont {F.}~\bibnamefont
  {Miao}}, \bibinfo {author} {\bibfnamefont {M.}~\bibnamefont {Lanza}},
  \bibinfo {author} {\bibfnamefont {T.~J.}\ \bibnamefont {Quill}}, \bibinfo
  {author} {\bibfnamefont {S.~T.}\ \bibnamefont {Keene}}, \bibinfo {author}
  {\bibfnamefont {A.}~\bibnamefont {Salleo}}, \bibinfo {author} {\bibfnamefont
  {J.}~\bibnamefont {Grollier}}, \bibinfo {author} {\bibfnamefont
  {D.}~\bibnamefont {Markovi{\'c}}}, \bibinfo {author} {\bibfnamefont
  {A.}~\bibnamefont {Mizrahi}}, \bibinfo {author} {\bibfnamefont
  {P.}~\bibnamefont {Yao}}, \bibinfo {author} {\bibfnamefont {J.~J.}\
  \bibnamefont {Yang}}, \bibinfo {author} {\bibfnamefont {G.}~\bibnamefont
  {Indiveri}}, \bibinfo {author} {\bibfnamefont {J.~P.}\ \bibnamefont
  {Strachan}}, \bibinfo {author} {\bibfnamefont {S.}~\bibnamefont {Datta}},
  \bibinfo {author} {\bibfnamefont {E.}~\bibnamefont {Vianello}}, \bibinfo
  {author} {\bibfnamefont {A.}~\bibnamefont {Valentian}}, \bibinfo {author}
  {\bibfnamefont {J.}~\bibnamefont {Feldmann}}, \bibinfo {author}
  {\bibfnamefont {X.}~\bibnamefont {Li}}, \bibinfo {author} {\bibfnamefont
  {W.~H.~P.}\ \bibnamefont {Pernice}}, \bibinfo {author} {\bibfnamefont
  {H.}~\bibnamefont {Bhaskaran}}, \bibinfo {author} {\bibfnamefont
  {S.}~\bibnamefont {Furber}}, \bibinfo {author} {\bibfnamefont
  {E.}~\bibnamefont {Neftci}}, \bibinfo {author} {\bibfnamefont
  {F.}~\bibnamefont {Scherr}}, \bibinfo {author} {\bibfnamefont
  {W.}~\bibnamefont {Maass}}, \bibinfo {author} {\bibfnamefont
  {S.}~\bibnamefont {Ramaswamy}}, \bibinfo {author} {\bibfnamefont
  {J.}~\bibnamefont {Tapson}}, \bibinfo {author} {\bibfnamefont
  {P.}~\bibnamefont {Panda}}, \bibinfo {author} {\bibfnamefont
  {Y.}~\bibnamefont {Kim}}, \bibinfo {author} {\bibfnamefont {G.}~\bibnamefont
  {Tanaka}}, \bibinfo {author} {\bibfnamefont {S.}~\bibnamefont {Thorpe}},
  \bibinfo {author} {\bibfnamefont {C.}~\bibnamefont {Bartolozzi}}, \bibinfo
  {author} {\bibfnamefont {T.~A.}\ \bibnamefont {Cleland}}, \bibinfo {author}
  {\bibfnamefont {C.}~\bibnamefont {Posch}}, \bibinfo {author} {\bibfnamefont
  {S.}~\bibnamefont {Liu}}, \bibinfo {author} {\bibfnamefont {G.}~\bibnamefont
  {Panuccio}}, \bibinfo {author} {\bibfnamefont {M.}~\bibnamefont {Mahmud}},
  \bibinfo {author} {\bibfnamefont {A.~N.}\ \bibnamefont {Mazumder}}, \bibinfo
  {author} {\bibfnamefont {M.}~\bibnamefont {Hosseini}}, \bibinfo {author}
  {\bibfnamefont {T.}~\bibnamefont {Mohsenin}}, \bibinfo {author}
  {\bibfnamefont {E.}~\bibnamefont {Donati}}, \bibinfo {author} {\bibfnamefont
  {S.}~\bibnamefont {Tolu}}, \bibinfo {author} {\bibfnamefont {R.}~\bibnamefont
  {Galeazzi}}, \bibinfo {author} {\bibfnamefont {M.~E.}\ \bibnamefont
  {Christensen}}, \bibinfo {author} {\bibfnamefont {S.}~\bibnamefont {Holm}},
  \bibinfo {author} {\bibfnamefont {D.}~\bibnamefont {Ielmini}},\ and\ \bibinfo
  {author} {\bibfnamefont {N.}~\bibnamefont {Pryds}},\ }\bibfield  {title}
  {\bibinfo {title} {2022 {{Roadmap}} on neuromorphic computing and
  engineering},\ }\href {https://doi.org/10.1088/2634-4386/ac4a83} {\bibfield
  {journal} {\bibinfo  {journal} {Neuromorphic Computing and Engineering}\
  }\textbf {\bibinfo {volume} {2}},\ \bibinfo {pages} {022501} (\bibinfo {year}
  {2022})}\BibitemShut {NoStop}%
\bibitem [{\citenamefont {Ziegler}\ \emph {et~al.}(2024)\citenamefont
  {Ziegler}, \citenamefont {Mussenbrock},\ and\ \citenamefont
  {Kohlstedt}}]{ziegler2024}%
  \BibitemOpen
  \bibinfo {editor} {\bibfnamefont {M.}~\bibnamefont {Ziegler}}, \bibinfo
  {editor} {\bibfnamefont {T.}~\bibnamefont {Mussenbrock}},\ and\ \bibinfo
  {editor} {\bibfnamefont {H.}~\bibnamefont {Kohlstedt}},\ eds.,\ \href
  {https://doi.org/10.1007/978-3-031-36705-2} {\emph {\bibinfo {title}
  {Bio-{{Inspired Information Pathways}}: {{From Neuroscience}} to
  {{Neurotronics}}}}},\ \bibinfo {series} {Springer {{Series}} on {{Bio-}} and
  {{Neurosystems}}}, Vol.~\bibinfo {volume} {16}\ (\bibinfo  {publisher}
  {Springer International Publishing},\ \bibinfo {address} {Cham},\ \bibinfo
  {year} {2024})\BibitemShut {NoStop}%
\bibitem [{\citenamefont {Ielmini}\ and\ \citenamefont
  {Wong}(2018)}]{ielmini2018}%
  \BibitemOpen
  \bibfield  {author} {\bibinfo {author} {\bibfnamefont {D.}~\bibnamefont
  {Ielmini}}\ and\ \bibinfo {author} {\bibfnamefont {H.-S.~P.}\ \bibnamefont
  {Wong}},\ }\bibfield  {title} {\bibinfo {title} {In-memory computing with
  resistive switching devices},\ }\href
  {https://doi.org/10.1038/s41928-018-0092-2} {\bibfield  {journal} {\bibinfo
  {journal} {Nature Electronics}\ }\textbf {\bibinfo {volume} {1}},\ \bibinfo
  {pages} {333} (\bibinfo {year} {2018})}\BibitemShut {NoStop}%
\bibitem [{\citenamefont {Ielmini}(2018)}]{ielmini2018a}%
  \BibitemOpen
  \bibfield  {author} {\bibinfo {author} {\bibfnamefont {D.}~\bibnamefont
  {Ielmini}},\ }\bibfield  {title} {\bibinfo {title} {Brain-inspired computing
  with resistive switching memory ({{RRAM}}): {{Devices}}, synapses and neural
  networks},\ }\href {https://doi.org/10.1016/j.mee.2018.01.009} {\bibfield
  {journal} {\bibinfo  {journal} {Microelectronic Engineering}\ }\textbf
  {\bibinfo {volume} {190}},\ \bibinfo {pages} {44} (\bibinfo {year}
  {2018})}\BibitemShut {NoStop}%
\bibitem [{\citenamefont {Ziegler}\ \emph {et~al.}(2018)\citenamefont
  {Ziegler}, \citenamefont {Wenger}, \citenamefont {Chicca},\ and\
  \citenamefont {Kohlstedt}}]{ziegler2018}%
  \BibitemOpen
  \bibfield  {author} {\bibinfo {author} {\bibfnamefont {M.}~\bibnamefont
  {Ziegler}}, \bibinfo {author} {\bibfnamefont {{\relax Ch}.}~\bibnamefont
  {Wenger}}, \bibinfo {author} {\bibfnamefont {E.}~\bibnamefont {Chicca}},\
  and\ \bibinfo {author} {\bibfnamefont {H.}~\bibnamefont {Kohlstedt}},\
  }\bibfield  {title} {\bibinfo {title} {Tutorial: {{Concepts}} for closely
  mimicking biological learning with memristive devices: {{Principles}} to
  emulate cellular forms of learning},\ }\href
  {https://doi.org/10.1063/1.5042040} {\bibfield  {journal} {\bibinfo
  {journal} {Journal of Applied Physics}\ }\textbf {\bibinfo {volume} {124}},\
  \bibinfo {pages} {152003} (\bibinfo {year} {2018})}\BibitemShut {NoStop}%
\bibitem [{\citenamefont {Ran}\ \emph {et~al.}(2023)\citenamefont {Ran},
  \citenamefont {Pei}, \citenamefont {Zhou}, \citenamefont {Wang},
  \citenamefont {Sun}, \citenamefont {Wang}, \citenamefont {Hao}, \citenamefont
  {Zhao}, \citenamefont {Chen},\ and\ \citenamefont {Yan}}]{ran2023}%
  \BibitemOpen
  \bibfield  {author} {\bibinfo {author} {\bibfnamefont {Y.}~\bibnamefont
  {Ran}}, \bibinfo {author} {\bibfnamefont {Y.}~\bibnamefont {Pei}}, \bibinfo
  {author} {\bibfnamefont {Z.}~\bibnamefont {Zhou}}, \bibinfo {author}
  {\bibfnamefont {H.}~\bibnamefont {Wang}}, \bibinfo {author} {\bibfnamefont
  {Y.}~\bibnamefont {Sun}}, \bibinfo {author} {\bibfnamefont {Z.}~\bibnamefont
  {Wang}}, \bibinfo {author} {\bibfnamefont {M.}~\bibnamefont {Hao}}, \bibinfo
  {author} {\bibfnamefont {J.}~\bibnamefont {Zhao}}, \bibinfo {author}
  {\bibfnamefont {J.}~\bibnamefont {Chen}},\ and\ \bibinfo {author}
  {\bibfnamefont {X.}~\bibnamefont {Yan}},\ }\bibfield  {title} {\bibinfo
  {title} {A review of {{Mott}} insulator in memristors: {{The}} materials,
  characteristics, applications for future computing systems and neuromorphic
  computing},\ }\href {https://doi.org/10.1007/s12274-022-4773-9} {\bibfield
  {journal} {\bibinfo  {journal} {Nano Research}\ }\textbf {\bibinfo {volume}
  {16}},\ \bibinfo {pages} {1165} (\bibinfo {year} {2023})}\BibitemShut
  {NoStop}%
\bibitem [{\citenamefont {Zhou}\ \emph {et~al.}(2022)\citenamefont {Zhou},
  \citenamefont {Wang}, \citenamefont {Sun}, \citenamefont {Zhou},
  \citenamefont {Sun}, \citenamefont {Zhao}, \citenamefont {Hu}, \citenamefont
  {Peng}, \citenamefont {Yan}, \citenamefont {Wang}, \citenamefont {Wang},
  \citenamefont {Li}, \citenamefont {Yan}, \citenamefont {Kuang}, \citenamefont
  {Wang}, \citenamefont {Wang},\ and\ \citenamefont {Duan}}]{zhou2022}%
  \BibitemOpen
  \bibfield  {author} {\bibinfo {author} {\bibfnamefont {G.}~\bibnamefont
  {Zhou}}, \bibinfo {author} {\bibfnamefont {Z.}~\bibnamefont {Wang}}, \bibinfo
  {author} {\bibfnamefont {B.}~\bibnamefont {Sun}}, \bibinfo {author}
  {\bibfnamefont {F.}~\bibnamefont {Zhou}}, \bibinfo {author} {\bibfnamefont
  {L.}~\bibnamefont {Sun}}, \bibinfo {author} {\bibfnamefont {H.}~\bibnamefont
  {Zhao}}, \bibinfo {author} {\bibfnamefont {X.}~\bibnamefont {Hu}}, \bibinfo
  {author} {\bibfnamefont {X.}~\bibnamefont {Peng}}, \bibinfo {author}
  {\bibfnamefont {J.}~\bibnamefont {Yan}}, \bibinfo {author} {\bibfnamefont
  {H.}~\bibnamefont {Wang}}, \bibinfo {author} {\bibfnamefont {W.}~\bibnamefont
  {Wang}}, \bibinfo {author} {\bibfnamefont {J.}~\bibnamefont {Li}}, \bibinfo
  {author} {\bibfnamefont {B.}~\bibnamefont {Yan}}, \bibinfo {author}
  {\bibfnamefont {D.}~\bibnamefont {Kuang}}, \bibinfo {author} {\bibfnamefont
  {Y.}~\bibnamefont {Wang}}, \bibinfo {author} {\bibfnamefont {L.}~\bibnamefont
  {Wang}},\ and\ \bibinfo {author} {\bibfnamefont {S.}~\bibnamefont {Duan}},\
  }\bibfield  {title} {\bibinfo {title} {Volatile and {{Nonvolatile Memristive
  Devices}} for {{Neuromorphic Computing}}},\ }\href
  {https://doi.org/10.1002/aelm.202101127} {\bibfield  {journal} {\bibinfo
  {journal} {Advanced Electronic Materials}\ }\textbf {\bibinfo {volume} {8}},\
  \bibinfo {pages} {2101127} (\bibinfo {year} {2022})}\BibitemShut {NoStop}%
\bibitem [{\citenamefont {Pickett}(2012)}]{pickett2012}%
  \BibitemOpen
  \bibfield  {author} {\bibinfo {author} {\bibfnamefont {M.~D.}\ \bibnamefont
  {Pickett}},\ }\bibfield  {title} {\bibinfo {title} {A scalable neuristor
  built with {{Mott}} memristors},\ }\href {https://doi.org/10.1038/nmat3510}
  {\bibfield  {journal} {\bibinfo  {journal} {Nature Materials}\ }\textbf
  {\bibinfo {volume} {12}},\ \bibinfo {pages} {114} (\bibinfo {year}
  {2012})}\BibitemShut {NoStop}%
\bibitem [{\citenamefont {Ignatov}\ \emph {et~al.}(2015)\citenamefont
  {Ignatov}, \citenamefont {Ziegler}, \citenamefont {Hansen}, \citenamefont
  {Petraru},\ and\ \citenamefont {Kohlstedt}}]{ignatov2015}%
  \BibitemOpen
  \bibfield  {author} {\bibinfo {author} {\bibfnamefont {M.}~\bibnamefont
  {Ignatov}}, \bibinfo {author} {\bibfnamefont {M.}~\bibnamefont {Ziegler}},
  \bibinfo {author} {\bibfnamefont {M.}~\bibnamefont {Hansen}}, \bibinfo
  {author} {\bibfnamefont {A.}~\bibnamefont {Petraru}},\ and\ \bibinfo {author}
  {\bibfnamefont {H.}~\bibnamefont {Kohlstedt}},\ }\bibfield  {title} {\bibinfo
  {title} {A memristive spiking neuron with firing rate coding},\ }\href
  {https://doi.org/10.3389/fnins.2015.00376} {\bibfield  {journal} {\bibinfo
  {journal} {Frontiers in Neuroscience}\ }\textbf {\bibinfo {volume} {9}},\
  \bibinfo {pages} {376} (\bibinfo {year} {2015})}\BibitemShut {NoStop}%
\bibitem [{\citenamefont {Stoliar}\ \emph {et~al.}(2017)\citenamefont
  {Stoliar}, \citenamefont {Tranchant}, \citenamefont {Corraze}, \citenamefont
  {Janod}, \citenamefont {Besland}, \citenamefont {Tesler}, \citenamefont
  {Rozenberg},\ and\ \citenamefont {Cario}}]{stoliar2017}%
  \BibitemOpen
  \bibfield  {author} {\bibinfo {author} {\bibfnamefont {P.}~\bibnamefont
  {Stoliar}}, \bibinfo {author} {\bibfnamefont {J.}~\bibnamefont {Tranchant}},
  \bibinfo {author} {\bibfnamefont {B.}~\bibnamefont {Corraze}}, \bibinfo
  {author} {\bibfnamefont {E.}~\bibnamefont {Janod}}, \bibinfo {author}
  {\bibfnamefont {M.~P.}\ \bibnamefont {Besland}}, \bibinfo {author}
  {\bibfnamefont {F.}~\bibnamefont {Tesler}}, \bibinfo {author} {\bibfnamefont
  {M.}~\bibnamefont {Rozenberg}},\ and\ \bibinfo {author} {\bibfnamefont
  {L.}~\bibnamefont {Cario}},\ }\bibfield  {title} {\bibinfo {title} {A
  {{Leaky}}-{{Integrate}}-and-{{Fire Neuron Analog Realized}} with a {{Mott
  Insulator}}},\ }\href {https://doi.org/10.1002/adfm.201604740} {\bibfield
  {journal} {\bibinfo  {journal} {Advanced Functional Materials}\ }\textbf
  {\bibinfo {volume} {27}},\ \bibinfo {pages} {1604740} (\bibinfo {year}
  {2017})}\BibitemShut {NoStop}%
\bibitem [{\citenamefont {Yi}\ \emph {et~al.}(2018)\citenamefont {Yi},
  \citenamefont {Tsang}, \citenamefont {Lam}, \citenamefont {Bai},
  \citenamefont {Crowell},\ and\ \citenamefont {Flores}}]{yi2018}%
  \BibitemOpen
  \bibfield  {author} {\bibinfo {author} {\bibfnamefont {W.}~\bibnamefont
  {Yi}}, \bibinfo {author} {\bibfnamefont {K.~K.}\ \bibnamefont {Tsang}},
  \bibinfo {author} {\bibfnamefont {S.~K.}\ \bibnamefont {Lam}}, \bibinfo
  {author} {\bibfnamefont {X.}~\bibnamefont {Bai}}, \bibinfo {author}
  {\bibfnamefont {J.~A.}\ \bibnamefont {Crowell}},\ and\ \bibinfo {author}
  {\bibfnamefont {E.~A.}\ \bibnamefont {Flores}},\ }\bibfield  {title}
  {\bibinfo {title} {Biological plausibility and stochasticity in scalable
  {{VO$_2$}} active memristor neurons},\ }\href
  {https://doi.org/10.1038/s41467-018-07052-w} {\bibfield  {journal} {\bibinfo
  {journal} {Nature Communications}\ }\textbf {\bibinfo {volume} {9}},\
  \bibinfo {pages} {4661} (\bibinfo {year} {2018})}\BibitemShut {NoStop}%
\bibitem [{\citenamefont {{del Valle}}\ \emph {et~al.}(2020)\citenamefont {{del
  Valle}}, \citenamefont {Salev}, \citenamefont {Kalcheim},\ and\ \citenamefont
  {Schuller}}]{delvalle2020}%
  \BibitemOpen
  \bibfield  {author} {\bibinfo {author} {\bibfnamefont {J.}~\bibnamefont {{del
  Valle}}}, \bibinfo {author} {\bibfnamefont {P.}~\bibnamefont {Salev}},
  \bibinfo {author} {\bibfnamefont {Y.}~\bibnamefont {Kalcheim}},\ and\
  \bibinfo {author} {\bibfnamefont {I.~K.}\ \bibnamefont {Schuller}},\
  }\bibfield  {title} {\bibinfo {title} {A caloritronics-based {{Mott}}
  neuristor},\ }\href {https://doi.org/10.1038/s41598-020-61176-y} {\bibfield
  {journal} {\bibinfo  {journal} {Scientific Reports}\ }\textbf {\bibinfo
  {volume} {10}},\ \bibinfo {pages} {4292} (\bibinfo {year}
  {2020})}\BibitemShut {NoStop}%
\bibitem [{\citenamefont {Ziegler}\ \emph {et~al.}(2015)\citenamefont
  {Ziegler}, \citenamefont {Riggert}, \citenamefont {Hansen}, \citenamefont
  {Bartsch},\ and\ \citenamefont {Kohlstedt}}]{ziegler2015}%
  \BibitemOpen
  \bibfield  {author} {\bibinfo {author} {\bibfnamefont {M.}~\bibnamefont
  {Ziegler}}, \bibinfo {author} {\bibfnamefont {C.}~\bibnamefont {Riggert}},
  \bibinfo {author} {\bibfnamefont {M.}~\bibnamefont {Hansen}}, \bibinfo
  {author} {\bibfnamefont {T.}~\bibnamefont {Bartsch}},\ and\ \bibinfo {author}
  {\bibfnamefont {H.}~\bibnamefont {Kohlstedt}},\ }\bibfield  {title} {\bibinfo
  {title} {Memristive {{Hebbian Plasticity Model}}: {{Device Requirements}} for
  the {{Emulation}} of {{Hebbian Plasticity Based}} on {{Memristive
  Devices}}},\ }\href {https://doi.org/10.1109/TBCAS.2015.2410811} {\bibfield
  {journal} {\bibinfo  {journal} {IEEE Transactions on Biomedical Circuits and
  Systems}\ }\textbf {\bibinfo {volume} {9}},\ \bibinfo {pages} {197} (\bibinfo
  {year} {2015})}\BibitemShut {NoStop}%
\bibitem [{\citenamefont {Hansen}\ \emph {et~al.}(2017)\citenamefont {Hansen},
  \citenamefont {Zahari}, \citenamefont {Ziegler},\ and\ \citenamefont
  {Kohlstedt}}]{hansen2017}%
  \BibitemOpen
  \bibfield  {author} {\bibinfo {author} {\bibfnamefont {M.}~\bibnamefont
  {Hansen}}, \bibinfo {author} {\bibfnamefont {F.}~\bibnamefont {Zahari}},
  \bibinfo {author} {\bibfnamefont {M.}~\bibnamefont {Ziegler}},\ and\ \bibinfo
  {author} {\bibfnamefont {H.}~\bibnamefont {Kohlstedt}},\ }\bibfield  {title}
  {\bibinfo {title} {Double-{{Barrier Memristive Devices}} for {{Unsupervised
  Learning}} and {{Pattern Recognition}}},\ }\href@noop {} {\bibfield
  {journal} {\bibinfo  {journal} {Frontiers in Neuroscience}\ }\textbf
  {\bibinfo {volume} {11}},\ \bibinfo {pages} {91} (\bibinfo {year}
  {2017})}\BibitemShut {NoStop}%
\bibitem [{\citenamefont {Burr}\ \emph {et~al.}(2017)\citenamefont {Burr},
  \citenamefont {Shelby}, \citenamefont {Sebastian}, \citenamefont {Kim},
  \citenamefont {Kim}, \citenamefont {Sidler}, \citenamefont {Virwani},
  \citenamefont {Ishii}, \citenamefont {Narayanan}, \citenamefont {Fumarola},
  \citenamefont {Sanches}, \citenamefont {Boybat}, \citenamefont {Le~Gallo},
  \citenamefont {Moon}, \citenamefont {Woo}, \citenamefont {Hwang},\ and\
  \citenamefont {Leblebici}}]{burr2017}%
  \BibitemOpen
  \bibfield  {author} {\bibinfo {author} {\bibfnamefont {G.~W.}\ \bibnamefont
  {Burr}}, \bibinfo {author} {\bibfnamefont {R.~M.}\ \bibnamefont {Shelby}},
  \bibinfo {author} {\bibfnamefont {A.}~\bibnamefont {Sebastian}}, \bibinfo
  {author} {\bibfnamefont {S.}~\bibnamefont {Kim}}, \bibinfo {author}
  {\bibfnamefont {S.}~\bibnamefont {Kim}}, \bibinfo {author} {\bibfnamefont
  {S.}~\bibnamefont {Sidler}}, \bibinfo {author} {\bibfnamefont
  {K.}~\bibnamefont {Virwani}}, \bibinfo {author} {\bibfnamefont
  {M.}~\bibnamefont {Ishii}}, \bibinfo {author} {\bibfnamefont
  {P.}~\bibnamefont {Narayanan}}, \bibinfo {author} {\bibfnamefont
  {A.}~\bibnamefont {Fumarola}}, \bibinfo {author} {\bibfnamefont {L.~L.}\
  \bibnamefont {Sanches}}, \bibinfo {author} {\bibfnamefont {I.}~\bibnamefont
  {Boybat}}, \bibinfo {author} {\bibfnamefont {M.}~\bibnamefont {Le~Gallo}},
  \bibinfo {author} {\bibfnamefont {K.}~\bibnamefont {Moon}}, \bibinfo {author}
  {\bibfnamefont {J.}~\bibnamefont {Woo}}, \bibinfo {author} {\bibfnamefont
  {H.}~\bibnamefont {Hwang}},\ and\ \bibinfo {author} {\bibfnamefont
  {Y.}~\bibnamefont {Leblebici}},\ }\bibfield  {title} {\bibinfo {title}
  {Neuromorphic computing using non-volatile memory},\ }\href
  {https://doi.org/10.1080/23746149.2016.1259585} {\bibfield  {journal}
  {\bibinfo  {journal} {Advances in Physics: X}\ }\textbf {\bibinfo {volume}
  {2}},\ \bibinfo {pages} {89} (\bibinfo {year} {2017})}\BibitemShut {NoStop}%
\bibitem [{\citenamefont {Jeong}\ and\ \citenamefont
  {Hwang}(2018)}]{jeong2018}%
  \BibitemOpen
  \bibfield  {author} {\bibinfo {author} {\bibfnamefont {D.~S.}\ \bibnamefont
  {Jeong}}\ and\ \bibinfo {author} {\bibfnamefont {C.~S.}\ \bibnamefont
  {Hwang}},\ }\bibfield  {title} {\bibinfo {title} {Nonvolatile {{Memory
  Materials}} for {{Neuromorphic Intelligent Machines}}},\ }\href
  {https://doi.org/10.1002/adma.201704729} {\bibfield  {journal} {\bibinfo
  {journal} {Advanced Materials}\ }\textbf {\bibinfo {volume} {30}},\ \bibinfo
  {pages} {1704729} (\bibinfo {year} {2018})}\BibitemShut {NoStop}%
\bibitem [{\citenamefont {del Valle}\ \emph {et~al.}(2018)\citenamefont {del
  Valle}, \citenamefont {Ramirez}, \citenamefont {Rozenberg},\ and\
  \citenamefont {Schuller}}]{valle2018}%
  \BibitemOpen
  \bibfield  {author} {\bibinfo {author} {\bibfnamefont {J.}~\bibnamefont {del
  Valle}}, \bibinfo {author} {\bibfnamefont {J.~G.}\ \bibnamefont {Ramirez}},
  \bibinfo {author} {\bibfnamefont {M.~J.}\ \bibnamefont {Rozenberg}},\ and\
  \bibinfo {author} {\bibfnamefont {I.~K.}\ \bibnamefont {Schuller}},\
  }\bibfield  {title} {\bibinfo {title} {Challenges in materials and devices
  for resistive-switching-based neuromorphic computing},\ }\href@noop {}
  {\bibfield  {journal} {\bibinfo  {journal} {Journal of Applied Physics}\
  }\textbf {\bibinfo {volume} {124}},\ \bibinfo {pages} {211101} (\bibinfo
  {year} {2018})}\BibitemShut {NoStop}%
\bibitem [{\citenamefont {Dittmann}\ \emph {et~al.}(2021)\citenamefont
  {Dittmann}, \citenamefont {Menzel},\ and\ \citenamefont
  {Waser}}]{dittmann2021}%
  \BibitemOpen
  \bibfield  {author} {\bibinfo {author} {\bibfnamefont {R.}~\bibnamefont
  {Dittmann}}, \bibinfo {author} {\bibfnamefont {S.}~\bibnamefont {Menzel}},\
  and\ \bibinfo {author} {\bibfnamefont {R.}~\bibnamefont {Waser}},\ }\bibfield
   {title} {\bibinfo {title} {Nanoionic memristive phenomena in metal oxides:
  The valence change mechanism},\ }\href
  {https://doi.org/10.1080/00018732.2022.2084006} {\bibfield  {journal}
  {\bibinfo  {journal} {Advances in Physics}\ }\textbf {\bibinfo {volume}
  {70}},\ \bibinfo {pages} {155} (\bibinfo {year} {2021})}\BibitemShut
  {NoStop}%
\bibitem [{\citenamefont {Hoffmann}\ \emph {et~al.}(2022)\citenamefont
  {Hoffmann}, \citenamefont {Ramanathan}, \citenamefont {Grollier},
  \citenamefont {Kent}, \citenamefont {Rozenberg}, \citenamefont {Schuller},
  \citenamefont {Shpyrko}, \citenamefont {Dynes}, \citenamefont {Fainman},
  \citenamefont {Frano}, \citenamefont {Fullerton}, \citenamefont {Galli},
  \citenamefont {Lomakin}, \citenamefont {Ong}, \citenamefont {{Petford-Long}},
  \citenamefont {Schuller}, \citenamefont {Stiles}, \citenamefont {Takamura},\
  and\ \citenamefont {Zhu}}]{hoffmann2022}%
  \BibitemOpen
  \bibfield  {author} {\bibinfo {author} {\bibfnamefont {A.}~\bibnamefont
  {Hoffmann}}, \bibinfo {author} {\bibfnamefont {S.}~\bibnamefont
  {Ramanathan}}, \bibinfo {author} {\bibfnamefont {J.}~\bibnamefont
  {Grollier}}, \bibinfo {author} {\bibfnamefont {A.~D.}\ \bibnamefont {Kent}},
  \bibinfo {author} {\bibfnamefont {M.~J.}\ \bibnamefont {Rozenberg}}, \bibinfo
  {author} {\bibfnamefont {I.~K.}\ \bibnamefont {Schuller}}, \bibinfo {author}
  {\bibfnamefont {O.~G.}\ \bibnamefont {Shpyrko}}, \bibinfo {author}
  {\bibfnamefont {R.~C.}\ \bibnamefont {Dynes}}, \bibinfo {author}
  {\bibfnamefont {Y.}~\bibnamefont {Fainman}}, \bibinfo {author} {\bibfnamefont
  {A.}~\bibnamefont {Frano}}, \bibinfo {author} {\bibfnamefont {E.~E.}\
  \bibnamefont {Fullerton}}, \bibinfo {author} {\bibfnamefont {G.}~\bibnamefont
  {Galli}}, \bibinfo {author} {\bibfnamefont {V.}~\bibnamefont {Lomakin}},
  \bibinfo {author} {\bibfnamefont {S.~P.}\ \bibnamefont {Ong}}, \bibinfo
  {author} {\bibfnamefont {A.~K.}\ \bibnamefont {{Petford-Long}}}, \bibinfo
  {author} {\bibfnamefont {J.~A.}\ \bibnamefont {Schuller}}, \bibinfo {author}
  {\bibfnamefont {M.~D.}\ \bibnamefont {Stiles}}, \bibinfo {author}
  {\bibfnamefont {Y.}~\bibnamefont {Takamura}},\ and\ \bibinfo {author}
  {\bibfnamefont {Y.}~\bibnamefont {Zhu}},\ }\bibfield  {title} {\bibinfo
  {title} {Quantum materials for energy-efficient neuromorphic computing:
  {{Opportunities}} and challenges},\ }\href
  {https://doi.org/10.1063/5.0094205} {\bibfield  {journal} {\bibinfo
  {journal} {APL Materials}\ }\textbf {\bibinfo {volume} {10}},\ \bibinfo
  {pages} {070904} (\bibinfo {year} {2022})}\BibitemShut {NoStop}%
\bibitem [{\citenamefont {Goteti}\ \emph {et~al.}(2021)\citenamefont {Goteti},
  \citenamefont {Zaluzhnyy}, \citenamefont {Ramanathan}, \citenamefont
  {Dynes},\ and\ \citenamefont {Frano}}]{goteti2021}%
  \BibitemOpen
  \bibfield  {author} {\bibinfo {author} {\bibfnamefont {U.~S.}\ \bibnamefont
  {Goteti}}, \bibinfo {author} {\bibfnamefont {I.~A.}\ \bibnamefont
  {Zaluzhnyy}}, \bibinfo {author} {\bibfnamefont {S.}~\bibnamefont
  {Ramanathan}}, \bibinfo {author} {\bibfnamefont {R.~C.}\ \bibnamefont
  {Dynes}},\ and\ \bibinfo {author} {\bibfnamefont {A.}~\bibnamefont {Frano}},\
  }\bibfield  {title} {\bibinfo {title} {Low-temperature emergent neuromorphic
  networks with correlated oxide devices},\ }\href
  {https://doi.org/10.1073/pnas.2103934118} {\bibfield  {journal} {\bibinfo
  {journal} {Proceedings of the National Academy of Sciences}\ }\textbf
  {\bibinfo {volume} {118}},\ \bibinfo {pages} {e2103934118} (\bibinfo {year}
  {2021})}\BibitemShut {NoStop}%
\bibitem [{\citenamefont {Imada}\ \emph {et~al.}(1998)\citenamefont {Imada},
  \citenamefont {Fujimori},\ and\ \citenamefont {Tokura}}]{imada1998}%
  \BibitemOpen
  \bibfield  {author} {\bibinfo {author} {\bibfnamefont {M.}~\bibnamefont
  {Imada}}, \bibinfo {author} {\bibfnamefont {A.}~\bibnamefont {Fujimori}},\
  and\ \bibinfo {author} {\bibfnamefont {Y.}~\bibnamefont {Tokura}},\
  }\bibfield  {title} {\bibinfo {title} {Metal-insulator transitions},\
  }\href@noop {} {\bibfield  {journal} {\bibinfo  {journal} {Reviews of Modern
  Physics}\ }\textbf {\bibinfo {volume} {70}},\ \bibinfo {pages} {1039}
  (\bibinfo {year} {1998})}\BibitemShut {NoStop}%
\bibitem [{\citenamefont {Catalano}\ \emph {et~al.}(2018)\citenamefont
  {Catalano}, \citenamefont {Gibert}, \citenamefont {Fowlie}, \citenamefont
  {{\'I}{\~n}iguez}, \citenamefont {Triscone},\ and\ \citenamefont
  {Kreisel}}]{catalano2018}%
  \BibitemOpen
  \bibfield  {author} {\bibinfo {author} {\bibfnamefont {S.}~\bibnamefont
  {Catalano}}, \bibinfo {author} {\bibfnamefont {M.}~\bibnamefont {Gibert}},
  \bibinfo {author} {\bibfnamefont {J.}~\bibnamefont {Fowlie}}, \bibinfo
  {author} {\bibfnamefont {J.}~\bibnamefont {{\'I}{\~n}iguez}}, \bibinfo
  {author} {\bibfnamefont {J.-M.}\ \bibnamefont {Triscone}},\ and\ \bibinfo
  {author} {\bibfnamefont {J.}~\bibnamefont {Kreisel}},\ }\bibfield  {title}
  {\bibinfo {title} {Rare-earth nickelates {{RNiO$_3$}}: Thin films and
  heterostructures},\ }\href {https://doi.org/10.1088/1361-6633/aaa37a}
  {\bibfield  {journal} {\bibinfo  {journal} {Reports on Progress in Physics}\
  }\textbf {\bibinfo {volume} {81}},\ \bibinfo {pages} {046501} (\bibinfo
  {year} {2018})}\BibitemShut {NoStop}%
\bibitem [{\citenamefont {Bisogni}\ \emph {et~al.}(2016)\citenamefont
  {Bisogni}, \citenamefont {Catalano}, \citenamefont {Green}, \citenamefont
  {Gibert}, \citenamefont {Scherwitzl}, \citenamefont {Huang}, \citenamefont
  {Strocov}, \citenamefont {Zubko}, \citenamefont {Balandeh}, \citenamefont
  {Triscone}, \citenamefont {Sawatzky},\ and\ \citenamefont
  {Schmitt}}]{bisogni2016}%
  \BibitemOpen
  \bibfield  {author} {\bibinfo {author} {\bibfnamefont {V.}~\bibnamefont
  {Bisogni}}, \bibinfo {author} {\bibfnamefont {S.}~\bibnamefont {Catalano}},
  \bibinfo {author} {\bibfnamefont {R.~J.}\ \bibnamefont {Green}}, \bibinfo
  {author} {\bibfnamefont {M.}~\bibnamefont {Gibert}}, \bibinfo {author}
  {\bibfnamefont {R.}~\bibnamefont {Scherwitzl}}, \bibinfo {author}
  {\bibfnamefont {Y.}~\bibnamefont {Huang}}, \bibinfo {author} {\bibfnamefont
  {V.~N.}\ \bibnamefont {Strocov}}, \bibinfo {author} {\bibfnamefont
  {P.}~\bibnamefont {Zubko}}, \bibinfo {author} {\bibfnamefont
  {S.}~\bibnamefont {Balandeh}}, \bibinfo {author} {\bibfnamefont {J.-M.}\
  \bibnamefont {Triscone}}, \bibinfo {author} {\bibfnamefont {G.}~\bibnamefont
  {Sawatzky}},\ and\ \bibinfo {author} {\bibfnamefont {T.}~\bibnamefont
  {Schmitt}},\ }\bibfield  {title} {\bibinfo {title} {Ground-state oxygen holes
  and the metal–insulator transition in the negative charge-transfer
  rare-earth nickelates},\ }\href {https://doi.org/10.1038/ncomms13017}
  {\bibfield  {journal} {\bibinfo  {journal} {Nature Communications}\ }\textbf
  {\bibinfo {volume} {7}},\ \bibinfo {pages} {13017} (\bibinfo {year}
  {2016})}\BibitemShut {NoStop}%
\bibitem [{\citenamefont {Johnston}\ \emph {et~al.}(2014)\citenamefont
  {Johnston}, \citenamefont {Mukherjee}, \citenamefont {Elfimov}, \citenamefont
  {Berciu},\ and\ \citenamefont {Sawatzky}}]{Johnston2014}%
  \BibitemOpen
  \bibfield  {author} {\bibinfo {author} {\bibfnamefont {S.}~\bibnamefont
  {Johnston}}, \bibinfo {author} {\bibfnamefont {A.}~\bibnamefont {Mukherjee}},
  \bibinfo {author} {\bibfnamefont {I.}~\bibnamefont {Elfimov}}, \bibinfo
  {author} {\bibfnamefont {M.}~\bibnamefont {Berciu}},\ and\ \bibinfo {author}
  {\bibfnamefont {G.~A.}\ \bibnamefont {Sawatzky}},\ }\bibfield  {title}
  {\bibinfo {title} {Charge {{Disproportionation}} without {{Charge Transfer}}
  in the {{Rare-Earth-Element Nickelates}} as a {{Possible Mechanism}} for the
  {{Metal-Insulator Transition}}},\ }\href
  {https://doi.org/10.1103/PhysRevLett.112.106404} {\bibfield  {journal}
  {\bibinfo  {journal} {Physical Review Letters}\ }\textbf {\bibinfo {volume}
  {112}},\ \bibinfo {pages} {106404} (\bibinfo {year} {2014})}\BibitemShut
  {NoStop}%
\bibitem [{\citenamefont {Berglund}(1969)}]{berglund1969}%
  \BibitemOpen
  \bibfield  {author} {\bibinfo {author} {\bibfnamefont {C.~N.}\ \bibnamefont
  {Berglund}},\ }\bibfield  {title} {\bibinfo {title} {Thermal filaments in
  vanadium dioxide},\ }\href {https://doi.org/10.1109/t-ed.1969.16773}
  {\bibfield  {journal} {\bibinfo  {journal} {IEEE Transactions on Electron
  Devices}\ }\textbf {\bibinfo {volume} {16}},\ \bibinfo {pages} {432}
  (\bibinfo {year} {1969})}\BibitemShut {NoStop}%
\bibitem [{\citenamefont {Zimmers}\ \emph {et~al.}(2013)\citenamefont
  {Zimmers}, \citenamefont {Aigouy}, \citenamefont {Mortier}, \citenamefont
  {Sharoni}, \citenamefont {Wang}, \citenamefont {West}, \citenamefont
  {Ram{\'i}rez},\ and\ \citenamefont {Schuller}}]{zimmers2013}%
  \BibitemOpen
  \bibfield  {author} {\bibinfo {author} {\bibfnamefont {A.}~\bibnamefont
  {Zimmers}}, \bibinfo {author} {\bibfnamefont {L.}~\bibnamefont {Aigouy}},
  \bibinfo {author} {\bibfnamefont {M.}~\bibnamefont {Mortier}}, \bibinfo
  {author} {\bibfnamefont {A.}~\bibnamefont {Sharoni}}, \bibinfo {author}
  {\bibfnamefont {S.}~\bibnamefont {Wang}}, \bibinfo {author} {\bibfnamefont
  {K.~G.}\ \bibnamefont {West}}, \bibinfo {author} {\bibfnamefont {J.~G.}\
  \bibnamefont {Ram{\'i}rez}},\ and\ \bibinfo {author} {\bibfnamefont {I.~K.}\
  \bibnamefont {Schuller}},\ }\bibfield  {title} {\bibinfo {title} {Role of
  {{Thermal Heating}} on the {{Voltage Induced Insulator-Metal Transition}} in
  {{VO$_2$}}},\ }\href {https://doi.org/10.1103/physrevlett.110.056601}
  {\bibfield  {journal} {\bibinfo  {journal} {Physical Review Letters}\
  }\textbf {\bibinfo {volume} {110}},\ \bibinfo {pages} {056601} (\bibinfo
  {year} {2013})}\BibitemShut {NoStop}%
\bibitem [{\citenamefont {Kumar}\ \emph {et~al.}(2013)\citenamefont {Kumar},
  \citenamefont {Pickett}, \citenamefont {Strachan}, \citenamefont {Gibson},
  \citenamefont {Nishi},\ and\ \citenamefont {Williams}}]{kumar2013}%
  \BibitemOpen
  \bibfield  {author} {\bibinfo {author} {\bibfnamefont {S.}~\bibnamefont
  {Kumar}}, \bibinfo {author} {\bibfnamefont {M.~D.}\ \bibnamefont {Pickett}},
  \bibinfo {author} {\bibfnamefont {J.~P.}\ \bibnamefont {Strachan}}, \bibinfo
  {author} {\bibfnamefont {G.}~\bibnamefont {Gibson}}, \bibinfo {author}
  {\bibfnamefont {Y.}~\bibnamefont {Nishi}},\ and\ \bibinfo {author}
  {\bibfnamefont {R.~S.}\ \bibnamefont {Williams}},\ }\bibfield  {title}
  {\bibinfo {title} {Local {{Temperature Redistribution}} and {{Structural
  Transition During Joule}}-{{Heating}}-{{Driven Conductance Switching}} in
  {{VO$_2$}}},\ }\href {https://doi.org/10.1002/adma.201302046} {\bibfield
  {journal} {\bibinfo  {journal} {Advanced Materials}\ }\textbf {\bibinfo
  {volume} {25}},\ \bibinfo {pages} {6128} (\bibinfo {year}
  {2013})}\BibitemShut {NoStop}%
\bibitem [{\citenamefont {Gu{\'e}non}\ \emph {et~al.}(2013)\citenamefont
  {Gu{\'e}non}, \citenamefont {Scharinger}, \citenamefont {Wang}, \citenamefont
  {Ram{\'i}rez}, \citenamefont {Koelle}, \citenamefont {Kleiner},\ and\
  \citenamefont {Schuller}}]{guenon2013}%
  \BibitemOpen
  \bibfield  {author} {\bibinfo {author} {\bibfnamefont {S.}~\bibnamefont
  {Gu{\'e}non}}, \bibinfo {author} {\bibfnamefont {S.}~\bibnamefont
  {Scharinger}}, \bibinfo {author} {\bibfnamefont {S.}~\bibnamefont {Wang}},
  \bibinfo {author} {\bibfnamefont {J.~G.}\ \bibnamefont {Ram{\'i}rez}},
  \bibinfo {author} {\bibfnamefont {D.}~\bibnamefont {Koelle}}, \bibinfo
  {author} {\bibfnamefont {R.}~\bibnamefont {Kleiner}},\ and\ \bibinfo {author}
  {\bibfnamefont {I.~K.}\ \bibnamefont {Schuller}},\ }\bibfield  {title}
  {\bibinfo {title} {Electrical breakdown in a {V$_2$O$_3$} device at the
  insulator-to-metal transition},\ }\href
  {https://doi.org/10.1209/0295-5075/101/57003} {\bibfield  {journal} {\bibinfo
   {journal} {Europhysics Letters (EPL)}\ }\textbf {\bibinfo {volume} {101}},\
  \bibinfo {pages} {57003} (\bibinfo {year} {2013})}\BibitemShut {NoStop}%
\bibitem [{\citenamefont {Manca}\ \emph {et~al.}(2015)\citenamefont {Manca},
  \citenamefont {Kanki}, \citenamefont {Tanaka}, \citenamefont {Marr{\'e}},\
  and\ \citenamefont {Pellegrino}}]{manca2015}%
  \BibitemOpen
  \bibfield  {author} {\bibinfo {author} {\bibfnamefont {N.}~\bibnamefont
  {Manca}}, \bibinfo {author} {\bibfnamefont {T.}~\bibnamefont {Kanki}},
  \bibinfo {author} {\bibfnamefont {H.}~\bibnamefont {Tanaka}}, \bibinfo
  {author} {\bibfnamefont {D.}~\bibnamefont {Marr{\'e}}},\ and\ \bibinfo
  {author} {\bibfnamefont {L.}~\bibnamefont {Pellegrino}},\ }\bibfield  {title}
  {\bibinfo {title} {Influence of thermal boundary conditions on the
  current-driven resistive transition in {{VO}}$_2$ microbridges},\ }\href@noop
  {} {\bibfield  {journal} {\bibinfo  {journal} {Applied Physics Letters}\
  }\textbf {\bibinfo {volume} {107}},\ \bibinfo {pages} {143509} (\bibinfo
  {year} {2015})}\BibitemShut {NoStop}%
\bibitem [{\citenamefont {Lange}\ \emph {et~al.}(2021)\citenamefont {Lange},
  \citenamefont {Gu{\'e}non}, \citenamefont {Kalcheim}, \citenamefont
  {Luibrand}, \citenamefont {Vargas}, \citenamefont {Schwebius}, \citenamefont
  {Kleiner}, \citenamefont {Schuller},\ and\ \citenamefont
  {Koelle}}]{lange2021}%
  \BibitemOpen
  \bibfield  {author} {\bibinfo {author} {\bibfnamefont {M.}~\bibnamefont
  {Lange}}, \bibinfo {author} {\bibfnamefont {S.}~\bibnamefont {Gu{\'e}non}},
  \bibinfo {author} {\bibfnamefont {Y.}~\bibnamefont {Kalcheim}}, \bibinfo
  {author} {\bibfnamefont {T.}~\bibnamefont {Luibrand}}, \bibinfo {author}
  {\bibfnamefont {N.~M.}\ \bibnamefont {Vargas}}, \bibinfo {author}
  {\bibfnamefont {D.}~\bibnamefont {Schwebius}}, \bibinfo {author}
  {\bibfnamefont {R.}~\bibnamefont {Kleiner}}, \bibinfo {author} {\bibfnamefont
  {I.~K.}\ \bibnamefont {Schuller}},\ and\ \bibinfo {author} {\bibfnamefont
  {D.}~\bibnamefont {Koelle}},\ }\bibfield  {title} {\bibinfo {title} {Imaging
  of {{Electrothermal Filament Formation}} in a {{Mott Insulator}}},\ }\href
  {https://doi.org/10.1103/physrevapplied.16.054027} {\bibfield  {journal}
  {\bibinfo  {journal} {Physical Review Applied}\ }\textbf {\bibinfo {volume}
  {16}},\ \bibinfo {pages} {054027} (\bibinfo {year} {2021})}\BibitemShut
  {NoStop}%
\bibitem [{\citenamefont {Rocco}\ \emph {et~al.}(2022)\citenamefont {Rocco},
  \citenamefont {del Valle}, \citenamefont {Navarro}, \citenamefont {Salev},
  \citenamefont {Schuller},\ and\ \citenamefont {Rozenberg}}]{rocco2022}%
  \BibitemOpen
  \bibfield  {author} {\bibinfo {author} {\bibfnamefont {R.}~\bibnamefont
  {Rocco}}, \bibinfo {author} {\bibfnamefont {J.}~\bibnamefont {del Valle}},
  \bibinfo {author} {\bibfnamefont {H.}~\bibnamefont {Navarro}}, \bibinfo
  {author} {\bibfnamefont {P.}~\bibnamefont {Salev}}, \bibinfo {author}
  {\bibfnamefont {I.~K.}\ \bibnamefont {Schuller}},\ and\ \bibinfo {author}
  {\bibfnamefont {M.}~\bibnamefont {Rozenberg}},\ }\bibfield  {title} {\bibinfo
  {title} {Exponential {{Escape Rate}} of {{Filamentary Incubation}} in {{Mott
  Spiking Neurons}}},\ }\href
  {https://doi.org/10.1103/physrevapplied.17.024028} {\bibfield  {journal}
  {\bibinfo  {journal} {Physical Review Applied}\ }\textbf {\bibinfo {volume}
  {17}},\ \bibinfo {pages} {024028} (\bibinfo {year} {2022})}\BibitemShut
  {NoStop}%
\bibitem [{\citenamefont {Luibrand}\ \emph {et~al.}(2023)\citenamefont
  {Luibrand}, \citenamefont {Bercher}, \citenamefont {Rocco}, \citenamefont
  {{Tahouni-Bonab}}, \citenamefont {Varbaro}, \citenamefont {Rischau},
  \citenamefont {Dominguez}, \citenamefont {Zhou}, \citenamefont {Luo},
  \citenamefont {Bag}, \citenamefont {Fratino}, \citenamefont {Kleiner},
  \citenamefont {Gariglio}, \citenamefont {Koelle}, \citenamefont {Triscone},
  \citenamefont {Rozenberg}, \citenamefont {Kuzmenko}, \citenamefont
  {Gu{\'e}non},\ and\ \citenamefont {{del Valle}}}]{luibrand2023}%
  \BibitemOpen
  \bibfield  {author} {\bibinfo {author} {\bibfnamefont {T.}~\bibnamefont
  {Luibrand}}, \bibinfo {author} {\bibfnamefont {A.}~\bibnamefont {Bercher}},
  \bibinfo {author} {\bibfnamefont {R.}~\bibnamefont {Rocco}}, \bibinfo
  {author} {\bibfnamefont {F.}~\bibnamefont {{Tahouni-Bonab}}}, \bibinfo
  {author} {\bibfnamefont {L.}~\bibnamefont {Varbaro}}, \bibinfo {author}
  {\bibfnamefont {C.~W.}\ \bibnamefont {Rischau}}, \bibinfo {author}
  {\bibfnamefont {C.}~\bibnamefont {Dominguez}}, \bibinfo {author}
  {\bibfnamefont {Y.}~\bibnamefont {Zhou}}, \bibinfo {author} {\bibfnamefont
  {W.}~\bibnamefont {Luo}}, \bibinfo {author} {\bibfnamefont {S.}~\bibnamefont
  {Bag}}, \bibinfo {author} {\bibfnamefont {L.}~\bibnamefont {Fratino}},
  \bibinfo {author} {\bibfnamefont {R.}~\bibnamefont {Kleiner}}, \bibinfo
  {author} {\bibfnamefont {S.}~\bibnamefont {Gariglio}}, \bibinfo {author}
  {\bibfnamefont {D.}~\bibnamefont {Koelle}}, \bibinfo {author} {\bibfnamefont
  {J.-M.}\ \bibnamefont {Triscone}}, \bibinfo {author} {\bibfnamefont {M.~J.}\
  \bibnamefont {Rozenberg}}, \bibinfo {author} {\bibfnamefont {A.~B.}\
  \bibnamefont {Kuzmenko}}, \bibinfo {author} {\bibfnamefont {S.}~\bibnamefont
  {Gu{\'e}non}},\ and\ \bibinfo {author} {\bibfnamefont {J.}~\bibnamefont {{del
  Valle}}},\ }\bibfield  {title} {\bibinfo {title} {Characteristic length
  scales of the electrically induced insulator-to-metal transition},\ }\href
  {https://doi.org/10.1103/PhysRevResearch.5.013108} {\bibfield  {journal}
  {\bibinfo  {journal} {Physical Review Research}\ }\textbf {\bibinfo {volume}
  {5}},\ \bibinfo {pages} {013108} (\bibinfo {year} {2023})}\BibitemShut
  {NoStop}%
\bibitem [{\citenamefont {del Valle}\ \emph {et~al.}(2019)\citenamefont {del
  Valle}, \citenamefont {Salev}, \citenamefont {Tesler}, \citenamefont
  {Vargas}, \citenamefont {Kalcheim}, \citenamefont {Wang}, \citenamefont
  {Trastoy}, \citenamefont {Lee}, \citenamefont {Kassabian}, \citenamefont
  {Ram{\'i}rez}, \citenamefont {Rozenberg},\ and\ \citenamefont
  {Schuller}}]{valle2019}%
  \BibitemOpen
  \bibfield  {author} {\bibinfo {author} {\bibfnamefont {J.}~\bibnamefont {del
  Valle}}, \bibinfo {author} {\bibfnamefont {P.}~\bibnamefont {Salev}},
  \bibinfo {author} {\bibfnamefont {F.}~\bibnamefont {Tesler}}, \bibinfo
  {author} {\bibfnamefont {N.~M.}\ \bibnamefont {Vargas}}, \bibinfo {author}
  {\bibfnamefont {Y.}~\bibnamefont {Kalcheim}}, \bibinfo {author}
  {\bibfnamefont {P.}~\bibnamefont {Wang}}, \bibinfo {author} {\bibfnamefont
  {J.}~\bibnamefont {Trastoy}}, \bibinfo {author} {\bibfnamefont {M.-H.}\
  \bibnamefont {Lee}}, \bibinfo {author} {\bibfnamefont {G.}~\bibnamefont
  {Kassabian}}, \bibinfo {author} {\bibfnamefont {J.~G.}\ \bibnamefont
  {Ram{\'i}rez}}, \bibinfo {author} {\bibfnamefont {M.~J.}\ \bibnamefont
  {Rozenberg}},\ and\ \bibinfo {author} {\bibfnamefont {I.~K.}\ \bibnamefont
  {Schuller}},\ }\bibfield  {title} {\bibinfo {title} {Subthreshold firing in
  {{Mott}} nanodevices},\ }\href {https://doi.org/10.1038/s41586-019-1159-6}
  {\bibfield  {journal} {\bibinfo  {journal} {Nature}\ }\textbf {\bibinfo
  {volume} {569}},\ \bibinfo {pages} {388} (\bibinfo {year}
  {2019})}\BibitemShut {NoStop}%
\bibitem [{\citenamefont {Adda}\ \emph {et~al.}(2022)\citenamefont {Adda},
  \citenamefont {Lee}, \citenamefont {Kalcheim}, \citenamefont {Salev},
  \citenamefont {Rocco}, \citenamefont {Vargas}, \citenamefont {Ghazikhanian},
  \citenamefont {Li}, \citenamefont {Albright}, \citenamefont {Rozenberg},\
  and\ \citenamefont {Schuller}}]{adda2022}%
  \BibitemOpen
  \bibfield  {author} {\bibinfo {author} {\bibfnamefont {C.}~\bibnamefont
  {Adda}}, \bibinfo {author} {\bibfnamefont {M.-H.}\ \bibnamefont {Lee}},
  \bibinfo {author} {\bibfnamefont {Y.}~\bibnamefont {Kalcheim}}, \bibinfo
  {author} {\bibfnamefont {P.}~\bibnamefont {Salev}}, \bibinfo {author}
  {\bibfnamefont {R.}~\bibnamefont {Rocco}}, \bibinfo {author} {\bibfnamefont
  {N.~M.}\ \bibnamefont {Vargas}}, \bibinfo {author} {\bibfnamefont
  {N.}~\bibnamefont {Ghazikhanian}}, \bibinfo {author} {\bibfnamefont {C.-P.}\
  \bibnamefont {Li}}, \bibinfo {author} {\bibfnamefont {G.}~\bibnamefont
  {Albright}}, \bibinfo {author} {\bibfnamefont {M.}~\bibnamefont
  {Rozenberg}},\ and\ \bibinfo {author} {\bibfnamefont {I.~K.}\ \bibnamefont
  {Schuller}},\ }\bibfield  {title} {\bibinfo {title} {Direct {{Observation}}
  of the {{Electrically Triggered Insulator-Metal Transition}} in {{V$_3$O$_5$
  Far}} below the {{Transition Temperature}}},\ }\href
  {https://doi.org/10.1103/physrevx.12.011025} {\bibfield  {journal} {\bibinfo
  {journal} {Physical Review X}\ }\textbf {\bibinfo {volume} {12}},\ \bibinfo
  {pages} {011025} (\bibinfo {year} {2022})}\BibitemShut {NoStop}%
\bibitem [{\citenamefont {Das}\ \emph {et~al.}(2023)\citenamefont {Das},
  \citenamefont {Nandi}, \citenamefont {Marquez}, \citenamefont {R{\'u}a},
  \citenamefont {Uenuma}, \citenamefont {Puyoo}, \citenamefont {Nath},
  \citenamefont {Albertini}, \citenamefont {Baboux}, \citenamefont {Lu},
  \citenamefont {Liu}, \citenamefont {Haeger}, \citenamefont {Heiderhoff},
  \citenamefont {Riedl}, \citenamefont {Ratcliff},\ and\ \citenamefont
  {Elliman}}]{das2023}%
  \BibitemOpen
  \bibfield  {author} {\bibinfo {author} {\bibfnamefont {S.~K.}\ \bibnamefont
  {Das}}, \bibinfo {author} {\bibfnamefont {S.~K.}\ \bibnamefont {Nandi}},
  \bibinfo {author} {\bibfnamefont {C.~V.}\ \bibnamefont {Marquez}}, \bibinfo
  {author} {\bibfnamefont {A.}~\bibnamefont {R{\'u}a}}, \bibinfo {author}
  {\bibfnamefont {M.}~\bibnamefont {Uenuma}}, \bibinfo {author} {\bibfnamefont
  {E.}~\bibnamefont {Puyoo}}, \bibinfo {author} {\bibfnamefont {S.~K.}\
  \bibnamefont {Nath}}, \bibinfo {author} {\bibfnamefont {D.}~\bibnamefont
  {Albertini}}, \bibinfo {author} {\bibfnamefont {N.}~\bibnamefont {Baboux}},
  \bibinfo {author} {\bibfnamefont {T.}~\bibnamefont {Lu}}, \bibinfo {author}
  {\bibfnamefont {Y.}~\bibnamefont {Liu}}, \bibinfo {author} {\bibfnamefont
  {T.}~\bibnamefont {Haeger}}, \bibinfo {author} {\bibfnamefont
  {R.}~\bibnamefont {Heiderhoff}}, \bibinfo {author} {\bibfnamefont
  {T.}~\bibnamefont {Riedl}}, \bibinfo {author} {\bibfnamefont
  {T.}~\bibnamefont {Ratcliff}},\ and\ \bibinfo {author} {\bibfnamefont
  {R.~G.}\ \bibnamefont {Elliman}},\ }\bibfield  {title} {\bibinfo {title}
  {Physical {{Origin}} of {{Negative Differential Resistance}} in {V$_3$O$_5$}
  and {{Its Application}} as a {{Solid-State Oscillator}}},\ }\href
  {https://doi.org/10.1002/adma.202208477} {\bibfield  {journal} {\bibinfo
  {journal} {Advanced Materials}\ }\textbf {\bibinfo {volume} {35}},\ \bibinfo
  {pages} {2208477} (\bibinfo {year} {2023})}\BibitemShut {NoStop}%
\bibitem [{\citenamefont {Catalan}(2008)}]{catalan2008}%
  \BibitemOpen
  \bibfield  {author} {\bibinfo {author} {\bibfnamefont {G.}~\bibnamefont
  {Catalan}},\ }\bibfield  {title} {\bibinfo {title} {Progress in perovskite
  nickelate research},\ }\href {https://doi.org/10.1080/01411590801992463}
  {\bibfield  {journal} {\bibinfo  {journal} {Phase Transitions}\ }\textbf
  {\bibinfo {volume} {81}},\ \bibinfo {pages} {729} (\bibinfo {year}
  {2008})}\BibitemShut {NoStop}%
\bibitem [{\citenamefont {Lu}\ \emph {et~al.}(2016)\citenamefont {Lu},
  \citenamefont {Frano}, \citenamefont {Bluschke}, \citenamefont {Hepting},
  \citenamefont {Macke}, \citenamefont {Strempfer}, \citenamefont {Wochner},
  \citenamefont {Cristiani}, \citenamefont {Logvenov}, \citenamefont
  {Habermeier}, \citenamefont {Haverkort}, \citenamefont {Keimer},\ and\
  \citenamefont {Benckiser}}]{lu2016}%
  \BibitemOpen
  \bibfield  {author} {\bibinfo {author} {\bibfnamefont {Y.}~\bibnamefont
  {Lu}}, \bibinfo {author} {\bibfnamefont {A.}~\bibnamefont {Frano}}, \bibinfo
  {author} {\bibfnamefont {M.}~\bibnamefont {Bluschke}}, \bibinfo {author}
  {\bibfnamefont {M.}~\bibnamefont {Hepting}}, \bibinfo {author} {\bibfnamefont
  {S.}~\bibnamefont {Macke}}, \bibinfo {author} {\bibfnamefont
  {J.}~\bibnamefont {Strempfer}}, \bibinfo {author} {\bibfnamefont
  {P.}~\bibnamefont {Wochner}}, \bibinfo {author} {\bibfnamefont
  {G.}~\bibnamefont {Cristiani}}, \bibinfo {author} {\bibfnamefont
  {G.}~\bibnamefont {Logvenov}}, \bibinfo {author} {\bibfnamefont {H.-U.}\
  \bibnamefont {Habermeier}}, \bibinfo {author} {\bibfnamefont {M.~W.}\
  \bibnamefont {Haverkort}}, \bibinfo {author} {\bibfnamefont {B.}~\bibnamefont
  {Keimer}},\ and\ \bibinfo {author} {\bibfnamefont {E.}~\bibnamefont
  {Benckiser}},\ }\bibfield  {title} {\bibinfo {title} {Quantitative
  determination of bond order and lattice distortions in nickel oxide
  heterostructures by resonant x-ray scattering},\ }\href
  {https://doi.org/10.1103/PhysRevB.93.165121} {\bibfield  {journal} {\bibinfo
  {journal} {Physical Review B}\ }\textbf {\bibinfo {volume} {93}},\ \bibinfo
  {pages} {165121} (\bibinfo {year} {2016})}\BibitemShut {NoStop}%
\bibitem [{\citenamefont {Liu}\ \emph {et~al.}(2013)\citenamefont {Liu},
  \citenamefont {Kargarian}, \citenamefont {Kareev}, \citenamefont {Gray},
  \citenamefont {Ryan}, \citenamefont {Cruz}, \citenamefont {Tahir},
  \citenamefont {Chuang}, \citenamefont {Guo}, \citenamefont {Rondinelli},
  \citenamefont {Freeland}, \citenamefont {Fiete},\ and\ \citenamefont
  {Chakhalian}}]{liu2013a}%
  \BibitemOpen
  \bibfield  {author} {\bibinfo {author} {\bibfnamefont {J.}~\bibnamefont
  {Liu}}, \bibinfo {author} {\bibfnamefont {M.}~\bibnamefont {Kargarian}},
  \bibinfo {author} {\bibfnamefont {M.}~\bibnamefont {Kareev}}, \bibinfo
  {author} {\bibfnamefont {B.}~\bibnamefont {Gray}}, \bibinfo {author}
  {\bibfnamefont {P.~J.}\ \bibnamefont {Ryan}}, \bibinfo {author}
  {\bibfnamefont {A.}~\bibnamefont {Cruz}}, \bibinfo {author} {\bibfnamefont
  {N.}~\bibnamefont {Tahir}}, \bibinfo {author} {\bibfnamefont {Y.-D.}\
  \bibnamefont {Chuang}}, \bibinfo {author} {\bibfnamefont {J.}~\bibnamefont
  {Guo}}, \bibinfo {author} {\bibfnamefont {J.~M.}\ \bibnamefont {Rondinelli}},
  \bibinfo {author} {\bibfnamefont {J.~W.}\ \bibnamefont {Freeland}}, \bibinfo
  {author} {\bibfnamefont {G.~A.}\ \bibnamefont {Fiete}},\ and\ \bibinfo
  {author} {\bibfnamefont {J.}~\bibnamefont {Chakhalian}},\ }\bibfield  {title}
  {\bibinfo {title} {Heterointerface engineered electronic and magnetic phases
  of {{NdNiO$_3$}} thin films},\ }\href {https://doi.org/10.1038/ncomms3714}
  {\bibfield  {journal} {\bibinfo  {journal} {Nature Communications}\ }\textbf
  {\bibinfo {volume} {4}},\ \bibinfo {pages} {2714} (\bibinfo {year}
  {2013})}\BibitemShut {NoStop}%
\bibitem [{\citenamefont {Catalano}\ \emph {et~al.}(2015)\citenamefont
  {Catalano}, \citenamefont {Gibert}, \citenamefont {Bisogni}, \citenamefont
  {He}, \citenamefont {Sutarto}, \citenamefont {Viret}, \citenamefont {Zubko},
  \citenamefont {Scherwitzl}, \citenamefont {Sawatzky}, \citenamefont
  {Schmitt},\ and\ \citenamefont {Triscone}}]{catalano2015}%
  \BibitemOpen
  \bibfield  {author} {\bibinfo {author} {\bibfnamefont {S.}~\bibnamefont
  {Catalano}}, \bibinfo {author} {\bibfnamefont {M.}~\bibnamefont {Gibert}},
  \bibinfo {author} {\bibfnamefont {V.}~\bibnamefont {Bisogni}}, \bibinfo
  {author} {\bibfnamefont {F.}~\bibnamefont {He}}, \bibinfo {author}
  {\bibfnamefont {R.}~\bibnamefont {Sutarto}}, \bibinfo {author} {\bibfnamefont
  {M.}~\bibnamefont {Viret}}, \bibinfo {author} {\bibfnamefont
  {P.}~\bibnamefont {Zubko}}, \bibinfo {author} {\bibfnamefont
  {R.}~\bibnamefont {Scherwitzl}}, \bibinfo {author} {\bibfnamefont {G.~A.}\
  \bibnamefont {Sawatzky}}, \bibinfo {author} {\bibfnamefont {T.}~\bibnamefont
  {Schmitt}},\ and\ \bibinfo {author} {\bibfnamefont {J.-M.}\ \bibnamefont
  {Triscone}},\ }\bibfield  {title} {\bibinfo {title} {Tailoring the electronic
  transitions of {{NdNiO$_3$}} films through {{(111)$_{\mathrm{pc}}$}} oriented
  interfaces},\ }\href {https://doi.org/10.1063/1.4919803} {\bibfield
  {journal} {\bibinfo  {journal} {APL Materials}\ }\textbf {\bibinfo {volume}
  {3}},\ \bibinfo {pages} {062506} (\bibinfo {year} {2015})}\BibitemShut
  {NoStop}%
\bibitem [{\citenamefont {Suyolcu}\ \emph {et~al.}(2021)\citenamefont
  {Suyolcu}, \citenamefont {Fürsich}, \citenamefont {Hepting}, \citenamefont
  {Zhong}, \citenamefont {Lu}, \citenamefont {Wang}, \citenamefont
  {Christiani}, \citenamefont {Logvenov}, \citenamefont {Hansmann},
  \citenamefont {Minola}, \citenamefont {Keimer}, \citenamefont {van Aken},\
  and\ \citenamefont {Benckiser}}]{suyolcu2021}%
  \BibitemOpen
  \bibfield  {author} {\bibinfo {author} {\bibfnamefont {Y.~E.}\ \bibnamefont
  {Suyolcu}}, \bibinfo {author} {\bibfnamefont {K.}~\bibnamefont {Fürsich}},
  \bibinfo {author} {\bibfnamefont {M.}~\bibnamefont {Hepting}}, \bibinfo
  {author} {\bibfnamefont {Z.}~\bibnamefont {Zhong}}, \bibinfo {author}
  {\bibfnamefont {Y.}~\bibnamefont {Lu}}, \bibinfo {author} {\bibfnamefont
  {Y.}~\bibnamefont {Wang}}, \bibinfo {author} {\bibfnamefont {G.}~\bibnamefont
  {Christiani}}, \bibinfo {author} {\bibfnamefont {G.}~\bibnamefont
  {Logvenov}}, \bibinfo {author} {\bibfnamefont {P.}~\bibnamefont {Hansmann}},
  \bibinfo {author} {\bibfnamefont {M.}~\bibnamefont {Minola}}, \bibinfo
  {author} {\bibfnamefont {B.}~\bibnamefont {Keimer}}, \bibinfo {author}
  {\bibfnamefont {P.~A.}\ \bibnamefont {van Aken}},\ and\ \bibinfo {author}
  {\bibfnamefont {E.}~\bibnamefont {Benckiser}},\ }\bibfield  {title} {\bibinfo
  {title} {Control of the metal-insulator transition in {$\mathrm{NdNiO}_{3}$}
  thin films through the interplay between structural and electronic
  properties},\ }\href {https://doi.org/10.1103/PhysRevMaterials.5.045001}
  {\bibfield  {journal} {\bibinfo  {journal} {Physical Review Materials}\
  }\textbf {\bibinfo {volume} {5}},\ \bibinfo {pages} {045001} (\bibinfo {year}
  {2021})}\BibitemShut {NoStop}%
\bibitem [{\citenamefont {Post}\ \emph {et~al.}(2018)\citenamefont {Post},
  \citenamefont {McLeod}, \citenamefont {Hepting}, \citenamefont {Bluschke},
  \citenamefont {Wang}, \citenamefont {Cristiani}, \citenamefont {Logvenov},
  \citenamefont {Charnukha}, \citenamefont {Ni}, \citenamefont {Radhakrishnan},
  \citenamefont {Minola}, \citenamefont {Pasupathy}, \citenamefont {Boris},
  \citenamefont {Benckiser}, \citenamefont {Dahmen}, \citenamefont {Carlson},
  \citenamefont {Keimer},\ and\ \citenamefont {Basov}}]{post2018}%
  \BibitemOpen
  \bibfield  {author} {\bibinfo {author} {\bibfnamefont {K.~W.}\ \bibnamefont
  {Post}}, \bibinfo {author} {\bibfnamefont {A.~S.}\ \bibnamefont {McLeod}},
  \bibinfo {author} {\bibfnamefont {M.}~\bibnamefont {Hepting}}, \bibinfo
  {author} {\bibfnamefont {M.}~\bibnamefont {Bluschke}}, \bibinfo {author}
  {\bibfnamefont {Y.}~\bibnamefont {Wang}}, \bibinfo {author} {\bibfnamefont
  {G.}~\bibnamefont {Cristiani}}, \bibinfo {author} {\bibfnamefont
  {G.}~\bibnamefont {Logvenov}}, \bibinfo {author} {\bibfnamefont
  {A.}~\bibnamefont {Charnukha}}, \bibinfo {author} {\bibfnamefont {G.~X.}\
  \bibnamefont {Ni}}, \bibinfo {author} {\bibfnamefont {P.}~\bibnamefont
  {Radhakrishnan}}, \bibinfo {author} {\bibfnamefont {M.}~\bibnamefont
  {Minola}}, \bibinfo {author} {\bibfnamefont {A.}~\bibnamefont {Pasupathy}},
  \bibinfo {author} {\bibfnamefont {A.~V.}\ \bibnamefont {Boris}}, \bibinfo
  {author} {\bibfnamefont {E.}~\bibnamefont {Benckiser}}, \bibinfo {author}
  {\bibfnamefont {K.~A.}\ \bibnamefont {Dahmen}}, \bibinfo {author}
  {\bibfnamefont {E.~W.}\ \bibnamefont {Carlson}}, \bibinfo {author}
  {\bibfnamefont {B.}~\bibnamefont {Keimer}},\ and\ \bibinfo {author}
  {\bibfnamefont {D.~N.}\ \bibnamefont {Basov}},\ }\bibfield  {title} {\bibinfo
  {title} {Coexisting first- and second-order electronic phase transitions in a
  correlated oxide},\ }\href {https://doi.org/10.1038/s41567-018-0201-1}
  {\bibfield  {journal} {\bibinfo  {journal} {Nature Physics}\ }\textbf
  {\bibinfo {volume} {14}},\ \bibinfo {pages} {1056} (\bibinfo {year}
  {2018})}\BibitemShut {NoStop}%
\bibitem [{\citenamefont {Lange}\ \emph {et~al.}(2017)\citenamefont {Lange},
  \citenamefont {Gu{\'e}non}, \citenamefont {Lever}, \citenamefont {Kleiner},\
  and\ \citenamefont {Koelle}}]{lange2017}%
  \BibitemOpen
  \bibfield  {author} {\bibinfo {author} {\bibfnamefont {M.~M.}\ \bibnamefont
  {Lange}}, \bibinfo {author} {\bibfnamefont {S.}~\bibnamefont {Gu{\'e}non}},
  \bibinfo {author} {\bibfnamefont {F.}~\bibnamefont {Lever}}, \bibinfo
  {author} {\bibfnamefont {R.}~\bibnamefont {Kleiner}},\ and\ \bibinfo {author}
  {\bibfnamefont {D.}~\bibnamefont {Koelle}},\ }\bibfield  {title} {\bibinfo
  {title} {A high-resolution combined scanning laser and widefield polarizing
  microscope for imaging at temperatures from 4\,{K} to 300\,{K}},\ }\href
  {https://doi.org/10.1063/1.5009529} {\bibfield  {journal} {\bibinfo
  {journal} {Review of Scientific Instruments}\ }\textbf {\bibinfo {volume}
  {88}},\ \bibinfo {pages} {123705} (\bibinfo {year} {2017})}\BibitemShut
  {NoStop}%
\bibitem [{\citenamefont {Lange}(2018)}]{lange2018}%
  \BibitemOpen
  \bibfield  {author} {\bibinfo {author} {\bibfnamefont {M.~M.}\ \bibnamefont
  {Lange}},\ }\emph {\bibinfo {title} {A {{High-Resolution Polarizing
  Microscope}} for {{Cryogenic Imaging}}: {{Development}} and {{Application}}
  to {{Investigations}} on {{Twin Walls}} in {\normalfont {{SrTiO}}$_3$} and
  the {{Metal-Insulator Transition}} in {\normalfont {V}$_2${{O}}$_3$}}},\
  \href@noop {} {Ph.D. thesis},\ \bibinfo  {school} {Universit{\"a}t
  T{\"u}bingen} (\bibinfo {year} {2018})\BibitemShut {NoStop}%
\bibitem [{\citenamefont {Rozenberg}\ \emph {et~al.}(2004)\citenamefont
  {Rozenberg}, \citenamefont {Inoue},\ and\ \citenamefont
  {S{\'a}nchez}}]{rozenberg2004}%
  \BibitemOpen
  \bibfield  {author} {\bibinfo {author} {\bibfnamefont {M.~J.}\ \bibnamefont
  {Rozenberg}}, \bibinfo {author} {\bibfnamefont {I.~H.}\ \bibnamefont
  {Inoue}},\ and\ \bibinfo {author} {\bibfnamefont {M.~J.}\ \bibnamefont
  {S{\'a}nchez}},\ }\bibfield  {title} {\bibinfo {title} {Nonvolatile
  {{Memory}} with {{Multilevel Switching}}: {{A Basic Model}}},\ }\href
  {https://doi.org/10.1103/PhysRevLett.92.178302} {\bibfield  {journal}
  {\bibinfo  {journal} {Physical Review Letters}\ }\textbf {\bibinfo {volume}
  {92}},\ \bibinfo {pages} {178302} (\bibinfo {year} {2004})}\BibitemShut
  {NoStop}%
\bibitem [{\citenamefont {Driscoll}\ \emph {et~al.}(2009)\citenamefont
  {Driscoll}, \citenamefont {Kim}, \citenamefont {Chae}, \citenamefont
  {Ventra},\ and\ \citenamefont {Basov}}]{driscoll2009}%
  \BibitemOpen
  \bibfield  {author} {\bibinfo {author} {\bibfnamefont {T.}~\bibnamefont
  {Driscoll}}, \bibinfo {author} {\bibfnamefont {H.-T.}\ \bibnamefont {Kim}},
  \bibinfo {author} {\bibfnamefont {B.-G.}\ \bibnamefont {Chae}}, \bibinfo
  {author} {\bibfnamefont {M.~D.}\ \bibnamefont {Ventra}},\ and\ \bibinfo
  {author} {\bibfnamefont {D.~N.}\ \bibnamefont {Basov}},\ }\bibfield  {title}
  {\bibinfo {title} {Phase-transition driven memristive system},\ }\href
  {https://doi.org/10.1063/1.3187531} {\bibfield  {journal} {\bibinfo
  {journal} {Applied Physics Letters}\ }\textbf {\bibinfo {volume} {95}},\
  \bibinfo {pages} {043503} (\bibinfo {year} {2009})}\BibitemShut {NoStop}%
\bibitem [{\citenamefont {Coy}\ \emph {et~al.}(2010)\citenamefont {Coy},
  \citenamefont {Cabrera}, \citenamefont {Sep{\'u}lveda},\ and\ \citenamefont
  {Fern{\'a}ndez}}]{coy2010}%
  \BibitemOpen
  \bibfield  {author} {\bibinfo {author} {\bibfnamefont {H.}~\bibnamefont
  {Coy}}, \bibinfo {author} {\bibfnamefont {R.}~\bibnamefont {Cabrera}},
  \bibinfo {author} {\bibfnamefont {N.}~\bibnamefont {Sep{\'u}lveda}},\ and\
  \bibinfo {author} {\bibfnamefont {F.~E.}\ \bibnamefont {Fern{\'a}ndez}},\
  }\bibfield  {title} {\bibinfo {title} {Optoelectronic and all-optical
  multiple memory states in vanadium dioxide},\ }\href
  {https://doi.org/10.1063/1.3518508} {\bibfield  {journal} {\bibinfo
  {journal} {Journal of Applied Physics}\ }\textbf {\bibinfo {volume} {108}},\
  \bibinfo {pages} {113115} (\bibinfo {year} {2010})}\BibitemShut {NoStop}%
\bibitem [{\citenamefont {Ha}\ \emph {et~al.}(2011)\citenamefont {Ha},
  \citenamefont {Aydogdu},\ and\ \citenamefont {Ramanathan}}]{ha2011}%
  \BibitemOpen
  \bibfield  {author} {\bibinfo {author} {\bibfnamefont {S.~D.}\ \bibnamefont
  {Ha}}, \bibinfo {author} {\bibfnamefont {G.~H.}\ \bibnamefont {Aydogdu}},\
  and\ \bibinfo {author} {\bibfnamefont {S.}~\bibnamefont {Ramanathan}},\
  }\bibfield  {title} {\bibinfo {title} {Metal-insulator transition and
  electrically driven memristive characteristics of {{SmNiO$_3$}} thin films},\
  }\href {https://doi.org/10.1063/1.3536486} {\bibfield  {journal} {\bibinfo
  {journal} {Applied Physics Letters}\ }\textbf {\bibinfo {volume} {98}},\
  \bibinfo {pages} {012105} (\bibinfo {year} {2011})}\BibitemShut {NoStop}%
\bibitem [{\citenamefont {Hu}\ \emph {et~al.}(2011)\citenamefont {Hu},
  \citenamefont {Zhang}, \citenamefont {Chen}, \citenamefont {Xu},\ and\
  \citenamefont {Wang}}]{hu2011}%
  \BibitemOpen
  \bibfield  {author} {\bibinfo {author} {\bibfnamefont {B.}~\bibnamefont
  {Hu}}, \bibinfo {author} {\bibfnamefont {Y.}~\bibnamefont {Zhang}}, \bibinfo
  {author} {\bibfnamefont {W.}~\bibnamefont {Chen}}, \bibinfo {author}
  {\bibfnamefont {C.}~\bibnamefont {Xu}},\ and\ \bibinfo {author}
  {\bibfnamefont {Z.~L.}\ \bibnamefont {Wang}},\ }\bibfield  {title} {\bibinfo
  {title} {Self-heating and {{External Strain Coupling Induced Phase
  Transition}} of {{VO$_2$}} {{Nanobeam}} as {{Single Domain Switch}}},\ }\href
  {https://doi.org/10.1002/adma.201101731} {\bibfield  {journal} {\bibinfo
  {journal} {Advanced Materials}\ }\textbf {\bibinfo {volume} {23}},\ \bibinfo
  {pages} {3536} (\bibinfo {year} {2011})}\BibitemShut {NoStop}%
\bibitem [{\citenamefont {Pellegrino}\ \emph {et~al.}(2012)\citenamefont
  {Pellegrino}, \citenamefont {Manca}, \citenamefont {Kanki}, \citenamefont
  {Tanaka}, \citenamefont {Biasotti}, \citenamefont {Bellingeri}, \citenamefont
  {Siri},\ and\ \citenamefont {Marr{\'e}}}]{pellegrino2012}%
  \BibitemOpen
  \bibfield  {author} {\bibinfo {author} {\bibfnamefont {L.}~\bibnamefont
  {Pellegrino}}, \bibinfo {author} {\bibfnamefont {N.}~\bibnamefont {Manca}},
  \bibinfo {author} {\bibfnamefont {T.}~\bibnamefont {Kanki}}, \bibinfo
  {author} {\bibfnamefont {H.}~\bibnamefont {Tanaka}}, \bibinfo {author}
  {\bibfnamefont {M.}~\bibnamefont {Biasotti}}, \bibinfo {author}
  {\bibfnamefont {E.}~\bibnamefont {Bellingeri}}, \bibinfo {author}
  {\bibfnamefont {A.~S.}\ \bibnamefont {Siri}},\ and\ \bibinfo {author}
  {\bibfnamefont {D.}~\bibnamefont {Marr{\'e}}},\ }\bibfield  {title} {\bibinfo
  {title} {Multistate {{Memory Devices Based}} on {{Free-standing
  VO}}$_2$/{{TiO}}$_2$ {{Microstructures Driven}} by {{Joule Self-Heating}}},\
  }\href {https://doi.org/10.1002/adma.201104669} {\bibfield  {journal}
  {\bibinfo  {journal} {Advanced Materials}\ }\textbf {\bibinfo {volume}
  {24}},\ \bibinfo {pages} {2929} (\bibinfo {year} {2012})}\BibitemShut
  {NoStop}%
\bibitem [{\citenamefont {Cai}\ \emph {et~al.}(2013)\citenamefont {Cai},
  \citenamefont {He}, \citenamefont {Sun}, \citenamefont {Li},\ and\
  \citenamefont {Zhou}}]{cai2013}%
  \BibitemOpen
  \bibfield  {author} {\bibinfo {author} {\bibfnamefont {K.}~\bibnamefont
  {Cai}}, \bibinfo {author} {\bibfnamefont {Z.}~\bibnamefont {He}}, \bibinfo
  {author} {\bibfnamefont {J.}~\bibnamefont {Sun}}, \bibinfo {author}
  {\bibfnamefont {B.}~\bibnamefont {Li}},\ and\ \bibinfo {author}
  {\bibfnamefont {J.}~\bibnamefont {Zhou}},\ }\bibfield  {title} {\bibinfo
  {title} {Resistive switching in a negative temperature coefficient metal
  oxide memristive one-port},\ }\href
  {https://doi.org/10.1007/s00339-012-7388-2} {\bibfield  {journal} {\bibinfo
  {journal} {Applied Physics A}\ }\textbf {\bibinfo {volume} {111}},\ \bibinfo
  {pages} {1045} (\bibinfo {year} {2013})}\BibitemShut {NoStop}%
\bibitem [{\citenamefont {Shi}\ \emph {et~al.}(2013)\citenamefont {Shi},
  \citenamefont {Ha}, \citenamefont {Zhou}, \citenamefont {Schoofs},\ and\
  \citenamefont {Ramanathan}}]{shi2013}%
  \BibitemOpen
  \bibfield  {author} {\bibinfo {author} {\bibfnamefont {J.}~\bibnamefont
  {Shi}}, \bibinfo {author} {\bibfnamefont {S.~D.}\ \bibnamefont {Ha}},
  \bibinfo {author} {\bibfnamefont {Y.}~\bibnamefont {Zhou}}, \bibinfo {author}
  {\bibfnamefont {F.}~\bibnamefont {Schoofs}},\ and\ \bibinfo {author}
  {\bibfnamefont {S.}~\bibnamefont {Ramanathan}},\ }\bibfield  {title}
  {\bibinfo {title} {A correlated nickelate synaptic transistor},\ }\href
  {https://doi.org/10.1038/ncomms3676} {\bibfield  {journal} {\bibinfo
  {journal} {Nature Communications}\ }\textbf {\bibinfo {volume} {4}},\
  \bibinfo {pages} {2676} (\bibinfo {year} {2013})}\BibitemShut {NoStop}%
\bibitem [{\citenamefont {Seo}\ \emph {et~al.}(2014)\citenamefont {Seo},
  \citenamefont {Kim}, \citenamefont {Kim},\ and\ \citenamefont
  {Lee}}]{seo2014}%
  \BibitemOpen
  \bibfield  {author} {\bibinfo {author} {\bibfnamefont {G.}~\bibnamefont
  {Seo}}, \bibinfo {author} {\bibfnamefont {B.-J.}\ \bibnamefont {Kim}},
  \bibinfo {author} {\bibfnamefont {H.-T.}\ \bibnamefont {Kim}},\ and\ \bibinfo
  {author} {\bibfnamefont {Y.~W.}\ \bibnamefont {Lee}},\ }\bibfield  {title}
  {\bibinfo {title} {Thermally- or optically-biased memristive switching in
  two-terminal {{VO$_2$}} devices},\ }\href
  {https://doi.org/10.1016/j.cap.2014.06.015} {\bibfield  {journal} {\bibinfo
  {journal} {Current Applied Physics}\ }\textbf {\bibinfo {volume} {14}},\
  \bibinfo {pages} {1251} (\bibinfo {year} {2014})}\BibitemShut {NoStop}%
\bibitem [{\citenamefont {Patel}\ \emph {et~al.}(2020)\citenamefont {Patel},
  \citenamefont {Okamoto}, \citenamefont {Dekker}, \citenamefont {Yang},
  \citenamefont {Gao}, \citenamefont {Luo}, \citenamefont {Lu}, \citenamefont
  {Sun},\ and\ \citenamefont {Tsen}}]{patel2020}%
  \BibitemOpen
  \bibfield  {author} {\bibinfo {author} {\bibfnamefont {T.}~\bibnamefont
  {Patel}}, \bibinfo {author} {\bibfnamefont {J.}~\bibnamefont {Okamoto}},
  \bibinfo {author} {\bibfnamefont {T.}~\bibnamefont {Dekker}}, \bibinfo
  {author} {\bibfnamefont {B.}~\bibnamefont {Yang}}, \bibinfo {author}
  {\bibfnamefont {J.}~\bibnamefont {Gao}}, \bibinfo {author} {\bibfnamefont
  {X.}~\bibnamefont {Luo}}, \bibinfo {author} {\bibfnamefont {W.}~\bibnamefont
  {Lu}}, \bibinfo {author} {\bibfnamefont {Y.}~\bibnamefont {Sun}},\ and\
  \bibinfo {author} {\bibfnamefont {A.~W.}\ \bibnamefont {Tsen}},\ }\bibfield
  {title} {\bibinfo {title} {Photocurrent {{Imaging}} of {{Multi-Memristive
  Charge Density Wave Switching}} in {{Two-Dimensional 1T-TaS}}$_2$},\ }\href
  {https://doi.org/10.1021/acs.nanolett.0c02537} {\bibfield  {journal}
  {\bibinfo  {journal} {Nano Letters}\ }\textbf {\bibinfo {volume} {20}},\
  \bibinfo {pages} {7200} (\bibinfo {year} {2020})}\BibitemShut {NoStop}%
\bibitem [{\citenamefont {Shabalin}\ \emph {et~al.}(2020)\citenamefont
  {Shabalin}, \citenamefont {{del Valle}}, \citenamefont {Hua}, \citenamefont
  {Cherukara}, \citenamefont {Holt}, \citenamefont {Schuller},\ and\
  \citenamefont {Shpyrko}}]{shabalin2020}%
  \BibitemOpen
  \bibfield  {author} {\bibinfo {author} {\bibfnamefont {A.~G.}\ \bibnamefont
  {Shabalin}}, \bibinfo {author} {\bibfnamefont {J.}~\bibnamefont {{del
  Valle}}}, \bibinfo {author} {\bibfnamefont {N.}~\bibnamefont {Hua}}, \bibinfo
  {author} {\bibfnamefont {M.~J.}\ \bibnamefont {Cherukara}}, \bibinfo {author}
  {\bibfnamefont {M.~V.}\ \bibnamefont {Holt}}, \bibinfo {author}
  {\bibfnamefont {I.~K.}\ \bibnamefont {Schuller}},\ and\ \bibinfo {author}
  {\bibfnamefont {O.~G.}\ \bibnamefont {Shpyrko}},\ }\bibfield  {title}
  {\bibinfo {title} {Nanoscale {{Imaging}} and {{Control}} of {{Volatile}} and
  {{Non-Volatile Resistive Switching}} in {{VO$_2$}}},\ }\href
  {https://doi.org/10.1002/smll.202005439} {\bibfield  {journal} {\bibinfo
  {journal} {Small}\ }\textbf {\bibinfo {volume} {16}},\ \bibinfo {pages}
  {2005439} (\bibinfo {year} {2020})}\BibitemShut {NoStop}%
\bibitem [{\citenamefont {Ma}\ \emph {et~al.}(2021)\citenamefont {Ma},
  \citenamefont {Wu}, \citenamefont {Li}, \citenamefont {Chen},\ and\
  \citenamefont {Lu}}]{ma2021}%
  \BibitemOpen
  \bibfield  {author} {\bibinfo {author} {\bibfnamefont {Y.}~\bibnamefont
  {Ma}}, \bibinfo {author} {\bibfnamefont {D.}~\bibnamefont {Wu}}, \bibinfo
  {author} {\bibfnamefont {Y.}~\bibnamefont {Li}}, \bibinfo {author}
  {\bibfnamefont {R.}~\bibnamefont {Chen}},\ and\ \bibinfo {author}
  {\bibfnamefont {C.}~\bibnamefont {Lu}},\ }\bibfield  {title} {\bibinfo
  {title} {Multiple {{Nonvolatile Resistance States Tuned}} by {{Electric
  Pulses}} in the {{Hysteresis Temperature Range}} of {{1T-TaS}}$_2$},\ }\href
  {https://doi.org/10.1002/andp.202000507} {\bibfield  {journal} {\bibinfo
  {journal} {Annalen der Physik}\ }\textbf {\bibinfo {volume} {533}},\ \bibinfo
  {pages} {2000507} (\bibinfo {year} {2021})}\BibitemShut {NoStop}%
\bibitem [{\citenamefont {Ma}\ \emph {et~al.}(2023)\citenamefont {Ma},
  \citenamefont {Xiao}, \citenamefont {Du}, \citenamefont {Li},\ and\
  \citenamefont {Wu}}]{ma2023a}%
  \BibitemOpen
  \bibfield  {author} {\bibinfo {author} {\bibfnamefont {Y.}~\bibnamefont
  {Ma}}, \bibinfo {author} {\bibfnamefont {G.}~\bibnamefont {Xiao}}, \bibinfo
  {author} {\bibfnamefont {L.}~\bibnamefont {Du}}, \bibinfo {author}
  {\bibfnamefont {A.}~\bibnamefont {Li}},\ and\ \bibinfo {author}
  {\bibfnamefont {D.}~\bibnamefont {Wu}},\ }\bibfield  {title} {\bibinfo
  {title} {Multiple resistance states induced by electric pulses and reset by
  joule heating in the hysteresis temperature range of
  {{(V$_{0.99}$Cr$_{0.01}$)$_2$O$_3$}}},\ }\href
  {https://doi.org/10.1063/5.0155698} {\bibfield  {journal} {\bibinfo
  {journal} {Applied Physics Letters}\ }\textbf {\bibinfo {volume} {123}},\
  \bibinfo {pages} {051901} (\bibinfo {year} {2023})}\BibitemShut {NoStop}%
\bibitem [{\citenamefont {Gurevich}\ and\ \citenamefont
  {Mints}(1987)}]{gurevich1987}%
  \BibitemOpen
  \bibfield  {author} {\bibinfo {author} {\bibfnamefont {A.~V.}\ \bibnamefont
  {Gurevich}}\ and\ \bibinfo {author} {\bibfnamefont {R.~G.}\ \bibnamefont
  {Mints}},\ }\bibfield  {title} {\bibinfo {title} {Self-heating in normal
  metals and superconductors},\ }\href
  {https://doi.org/10.1103/revmodphys.59.941} {\bibfield  {journal} {\bibinfo
  {journal} {Reviews of Modern Physics}\ }\textbf {\bibinfo {volume} {59}},\
  \bibinfo {pages} {941} (\bibinfo {year} {1987})}\BibitemShut {NoStop}%
\bibitem [{\citenamefont {Doenitz}\ \emph {et~al.}(2007)\citenamefont
  {Doenitz}, \citenamefont {Kleiner}, \citenamefont {Koelle}, \citenamefont
  {Scherer},\ and\ \citenamefont {Schuster}}]{doenitz2007}%
  \BibitemOpen
  \bibfield  {author} {\bibinfo {author} {\bibfnamefont {D.}~\bibnamefont
  {Doenitz}}, \bibinfo {author} {\bibfnamefont {R.}~\bibnamefont {Kleiner}},
  \bibinfo {author} {\bibfnamefont {D.}~\bibnamefont {Koelle}}, \bibinfo
  {author} {\bibfnamefont {T.}~\bibnamefont {Scherer}},\ and\ \bibinfo {author}
  {\bibfnamefont {K.~F.}\ \bibnamefont {Schuster}},\ }\bibfield  {title}
  {\bibinfo {title} {Imaging of thermal domains in ultrathin {{NbN}} films for
  hot electron bolometers},\ }\href {https://doi.org/10.1063/1.2751109}
  {\bibfield  {journal} {\bibinfo  {journal} {Applied Physics Letters}\
  }\textbf {\bibinfo {volume} {90}},\ \bibinfo {pages} {252512} (\bibinfo
  {year} {2007})}\BibitemShut {NoStop}%
\bibitem [{\citenamefont {Gross}\ \emph {et~al.}(2012)\citenamefont {Gross},
  \citenamefont {Gu{\'e}non}, \citenamefont {Yuan}, \citenamefont {Li},
  \citenamefont {Li}, \citenamefont {Ishii}, \citenamefont {Mints},
  \citenamefont {Hatano}, \citenamefont {Wu}, \citenamefont {Koelle},
  \citenamefont {Wang},\ and\ \citenamefont {Kleiner}}]{gross2012}%
  \BibitemOpen
  \bibfield  {author} {\bibinfo {author} {\bibfnamefont {B.}~\bibnamefont
  {Gross}}, \bibinfo {author} {\bibfnamefont {S.}~\bibnamefont {Gu{\'e}non}},
  \bibinfo {author} {\bibfnamefont {J.}~\bibnamefont {Yuan}}, \bibinfo {author}
  {\bibfnamefont {M.~Y.}\ \bibnamefont {Li}}, \bibinfo {author} {\bibfnamefont
  {J.}~\bibnamefont {Li}}, \bibinfo {author} {\bibfnamefont {A.}~\bibnamefont
  {Ishii}}, \bibinfo {author} {\bibfnamefont {R.~G.}\ \bibnamefont {Mints}},
  \bibinfo {author} {\bibfnamefont {T.}~\bibnamefont {Hatano}}, \bibinfo
  {author} {\bibfnamefont {P.~H.}\ \bibnamefont {Wu}}, \bibinfo {author}
  {\bibfnamefont {D.}~\bibnamefont {Koelle}}, \bibinfo {author} {\bibfnamefont
  {H.~B.}\ \bibnamefont {Wang}},\ and\ \bibinfo {author} {\bibfnamefont
  {R.}~\bibnamefont {Kleiner}},\ }\bibfield  {title} {\bibinfo {title}
  {Hot-spot formation in stacks of intrinsic {{Josephson}} junctions in
  {$\mathrm{Bi}_{2}\mathrm{Sr}_{2}\mathrm{CaCu}_{2}\mathrm{O}_{8}$}},\ }\href
  {https://doi.org/10.1103/physrevb.86.094524} {\bibfield  {journal} {\bibinfo
  {journal} {Physical Review B}\ }\textbf {\bibinfo {volume} {86}},\ \bibinfo
  {pages} {094524} (\bibinfo {year} {2012})}\BibitemShut {NoStop}%
\bibitem [{\citenamefont {B{\"u}ttiker}\ and\ \citenamefont
  {Landauer}(1982)}]{buttiker1982}%
  \BibitemOpen
  \bibfield  {author} {\bibinfo {author} {\bibfnamefont {M.}~\bibnamefont
  {B{\"u}ttiker}}\ and\ \bibinfo {author} {\bibfnamefont {R.}~\bibnamefont
  {Landauer}},\ }\bibfield  {title} {\bibinfo {title} {Transport and
  {{Fluctuations}} in {{Linear Arrays}} of {{Multistable Systems}}},\ }in\
  \href {https://doi.org/10.1007/978-1-4684-4127-7_8} {\emph {\bibinfo
  {booktitle} {{{NATO Science Series B}}: {{Nonlinear Phenomena}} at {{Phase
  Transitions}} and {{Instabilities}}}}},\ Vol.~\bibinfo {volume} {77},\
  \bibinfo {editor} {edited by\ \bibinfo {editor} {\bibfnamefont
  {T.}~\bibnamefont {Riste}}}\ (\bibinfo {year} {1982})\ pp.\ \bibinfo {pages}
  {111--143}\BibitemShut {NoStop}%
\bibitem [{\citenamefont {Spenke}(1936{\natexlab{a}})}]{spenke1936}%
  \BibitemOpen
  \bibfield  {author} {\bibinfo {author} {\bibfnamefont {E.}~\bibnamefont
  {Spenke}},\ }\bibfield  {title} {\bibinfo {title} {Zur technischen
  {{Beherrschung}} des {{W{\"a}rmedurchschlages}} von {{Hei{\ss}leitern}}},\
  }\href@noop {} {\bibfield  {journal} {\bibinfo  {journal} {Wissenschaftliche
  Ver{\"o}ffentlichungen aus den Siemens-Werken}\ }\textbf {\bibinfo {volume}
  {15}},\ \bibinfo {pages} {92} (\bibinfo {year}
  {1936}{\natexlab{a}})}\BibitemShut {NoStop}%
\bibitem [{\citenamefont {Spenke}(1936{\natexlab{b}})}]{spenke1936a}%
  \BibitemOpen
  \bibfield  {author} {\bibinfo {author} {\bibfnamefont {E.}~\bibnamefont
  {Spenke}},\ }\bibfield  {title} {\bibinfo {title} {Eine anschauliche
  {{Deutung}} der {{Abzweigtemperatur}} scheibenf{\"o}rmiger
  {{Hei{\ss}leiter}}.},\ }\href {https://doi.org/10.1007/bf01660663} {\bibfield
   {journal} {\bibinfo  {journal} {Archiv f{\"u}r Elektrotechnik}\ }\textbf
  {\bibinfo {volume} {30}},\ \bibinfo {pages} {728} (\bibinfo {year}
  {1936}{\natexlab{b}})}\BibitemShut {NoStop}%
\bibitem [{\citenamefont {Ridley}(1963)}]{ridley1963}%
  \BibitemOpen
  \bibfield  {author} {\bibinfo {author} {\bibfnamefont {B.~K.}\ \bibnamefont
  {Ridley}},\ }\bibfield  {title} {\bibinfo {title} {Specific {{Negative
  Resistance}} in {{Solids}}},\ }\href
  {https://doi.org/10.1088/0370-1328/82/6/315} {\bibfield  {journal} {\bibinfo
  {journal} {Proceedings of the Physical Society}\ }\textbf {\bibinfo {volume}
  {82}},\ \bibinfo {pages} {954} (\bibinfo {year} {1963})}\BibitemShut
  {NoStop}%
\end{thebibliography}%

\end{document}